\documentclass[acmtog,authorversion]{acmart}
\setcitestyle{square}

\usepackage{bm}
\usepackage{amsfonts}
\usepackage{amssymb}
\usepackage{algorithmic}
\usepackage{amsmath}
\usepackage{graphicx}
\usepackage{caption}
\usepackage{subfigure}
\usepackage{array}
\usepackage{enumitem}
\usepackage{multirow}
\usepackage{xspace}
\usepackage{gensymb}

\usepackage[ruled]{algorithm2e} 

\SetAlFnt{\small}
\SetAlCapFnt{\small}
\SetAlCapNameFnt{\small}
\SetAlCapHSkip{0pt}
\IncMargin{-\parindent}

\author{Xue Bin Peng}
\affiliation{\institution{University of California, Berkeley}}
\author{Angjoo Kanazawa}
\affiliation{\institution{University of California, Berkeley}}
\author{Jitendra Malik}
\affiliation{\institution{University of California, Berkeley}}
\author{Pieter Abbeel}
\affiliation{\institution{University of California, Berkeley}}
\author{Sergey Levine}
\affiliation{\institution{University of California, Berkeley}}


\title{SFV: Reinforcement Learning of Physical Skills from Videos}

\begin{document}

\setcopyright{acmlicensed}
\acmJournal{TOG}
\acmYear{2018}
\acmVolume{37}
\acmNumber{6}
\acmMonth{11}
\acmArticle{178} 
\acmDOI{10.1145/3272127.3275014}

\begin{teaserfigure}
 \centering
\includegraphics[width=\textwidth]{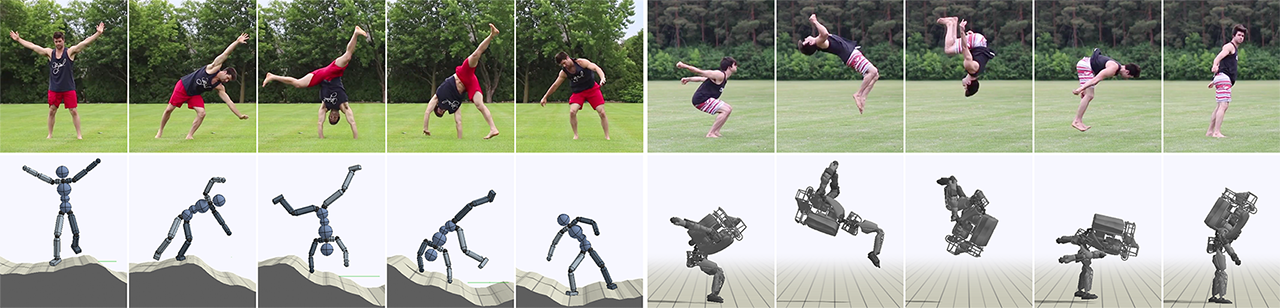}
   \caption{Simulated characters performing highly dynamic skills learned by imitating video clips of human demonstrations. \textbf{Left:} Humanoid performing cartwheel B on irregular terrain. \textbf{Right:} Backflip A retargeted to a simulated Atlas robot.}
\end{teaserfigure}

\begin{abstract}
Data-driven character animation based on motion capture can produce highly naturalistic behaviors and, when combined with physics simulation, can provide for natural procedural responses to physical perturbations, environmental changes, and morphological discrepancies. Motion capture remains the most popular source of motion data, but collecting mocap data typically requires heavily instrumented environments and actors. In this paper, we propose a method that enables physically simulated characters to learn skills from videos (SFV). Our approach, based on deep pose estimation and deep reinforcement learning, allows data-driven animation to leverage the abundance of publicly available video clips from the web, such as those from YouTube. This has the potential to enable fast and easy design of character controllers simply by querying for video recordings of the desired behavior. The resulting controllers are robust to perturbations, can be adapted to new settings, can perform basic object interactions, and can be retargeted to new morphologies via reinforcement learning.
We further demonstrate that our method can predict potential human motions from still images, by forward simulation of learned controllers initialized from the observed pose. 
Our framework is able to learn a broad range of dynamic skills, including locomotion, acrobatics, and martial arts. \href{https://xbpeng.github.io/projects/SFV/index.html}{(Video\footnote{\url{https://xbpeng.github.io/projects/SFV/index.html}})}
\end{abstract}


\maketitle

\section{Introduction}

Data-driven methods have been a cornerstone of character animation for decades, with motion-capture being one of the most popular sources of motion data. Mocap data is a staple for kinematic methods, and is also widely used in physics-based character animation. Imitation of mocap clips has been shown to be an effective approach for developing controllers for simulated characters, yielding some of the most diverse and naturalistic behaviors. However, the acquisition of mocap data can pose major hurdles for practitioners, often requiring heavily instrumented environments and actors. The infrastructure required to procure such data can be prohibitive, and some activities remain exceedingly difficult to motion capture, such as large-scale outdoor sports. A more abundant and flexible source of motion data is monocular video. A staggering 300 hours of video is uploaded to YouTube every minute \cite{YouTubeStats}. Searching and querying video sources on the web can quickly yield a large number of clips for any desired activity or behavior. However, it is a daunting challenge to extract the necessary motion information from monocular video frames, and the quality of the motions generated by previous methods still falls well behind the best mocap-based animation systems \cite{Vondrak2012}.

In this paper, we propose a method for acquiring dynamic character controllers directly from monocular video through a combination of pose estimation and deep reinforcement learning. Recent advances with deep learning techniques have produced breakthrough results for vision-based 3D pose estimation from monocular images \cite{hmrKanazawa17}.
However, pose estimation alone is not yet sufficient to produce high-fidelity and physically plausible motions: frequent errors and physical inconsistencies in the estimated poses accumulate and result in unnatural character behaviors. Motion imitation with reinforcement learning provides a powerful tool for acquiring skills from videos while remaining robust to such imperfections. By reproducing the skill in a physical simulation, the learning process can refine imperfect and noisy pose sequences, compensate for missing frames, and take into account the physical constraints of the character and environment. By bringing together deep pose estimation and reinforcement learning, we propose a framework that enables simulated characters to learn a diverse collection of dynamic and acrobatic skills directly from video demonstrations.

The primary contribution of our paper is a system for learning character controllers from video clips that integrates pose estimation and reinforcement learning. To make this possible, we introduce a number of extensions to both the pose tracking system and the reinforcement learning algorithm. We propose a motion reconstruction method that improves the quality of reference motions to be more amenable for imitation by a simulated character. We further introduce a novel reinforcement learning method that incorporates \emph{adaptive} state initialization, where the initial state distribution is dynamically updated to facilitate long-horizon performance in reproducing a desired motion. We find that this approach for dynamic curriculum generation substantially outperforms standard methods when learning from lower-fidelity reference motions constructed from video tracking sequences. Our framework is able to reproduce a significantly larger repertoire of skills and higher fidelity motions from videos than has been demonstrated by prior methods. The effectiveness of our framework is evaluated on a large set of challenging skills including dances, acrobatics, and martial arts. Our system is also able to retarget video demonstrations to widely different morphologies and environments. Figure 1 illustrates examples of the skills learned by our framework. Furthermore, we demonstrate a novel physics-based motion completion application that leverages a corpus of learned controllers to predict an actor's full-body motion given a single still image. While our framework is able to reproduce a substantially larger corpus of skills than previous methods, there remains a large variety of video clips that our system is not yet able to imitate. We include a discussion of these challenges and other limitations that arise from the various design decisions.

\section{Related Work}

Our work lies at the intersection of pose estimation and physics-based character animation. The end goal of our system is to produce robust and naturalistic controllers that enable virtual characters to perform complex skills in physically simulated environments. Facets of this problem have been studied in a large body of prior work, from techniques that have sought to produce realistic skills from first principles (i.e. physics and biomechanics) \cite{2010-TOG-gbwc,Wang2012,Wampler2014}, to methods that incorporate reference motion data into the controller construction process~\cite{daSilva2008a,Lee2010,2010-TOG-sampControl}. These techniques can synthesize motions kinematically \cite{Lee2010MotionFields,2012-ccclde,Holden2017} or as the product of dynamic control in a physics simulation~\cite{Lee2014,2013-TOG-MuscleBasedBipeds}. Most data-driven methods, save for a few exceptions, are based on motion capture data, which often requires costly instrumentation and pre-processing \cite{Holden2016}. Raw video offers a potentially more accessible and abundant alternative source of motion data. While there has been much progress in the computer vision community in predicting human poses from monocular images or videos, integrating pose predictions from video with data-driven character animation still presents a number of challenges. Pose estimators can generally produce reasonable predictions of an actor's motion, but they do not benefit from the manual cleanup and accurate tracking enjoyed by professionally recorded mocap data. Prior methods that learn from motion data often assume accurate reference motions as a vital component in the learning process. For example, during training, \cite{2018-TOG-DeepMimic} reinitializes the character state to frames sampled from the reference motion. The effectiveness of these strategies tend to deteriorate in the presence of low-fidelity reference motions.

\paragraph{Reinforcement Learning:} Many methods for acquiring character controllers utilize reinforcement learning \cite{Coros09,Wang2010,Lee2010MotionFields,2012-ccclde,2015-TOG-terrainRL}. The use of deep neural network models for RL has been demonstrated for a diverse array of challenging skills \cite{DuanCHSA16,BrockmanCPSSTZ16,2016-TOG-deepRL,Liu2017,Rajeswaran2017,TehBCQKHHP17}. While deep RL methods have been effective for motion control tasks, the policies are prone to developing unnatural behaviours, such as awkward postures, overly energetic movements, and asymmetric gaits \cite{SchulmanMLJA15,MerelTTSLWWH17}. In order to mitigate these artifacts, additional auxiliary objectives such as symmetry, effort minimization, or impact penalties have been incorporated into the objective to discourage unnatural behaviors \cite{YuMSL2018}. Designing effective objectives can require substantial human insight and may nonetheless fall short of eliminating undesirable behaviours. An alternative for encouraging more natural motions is to incorporate high-fidelity biomechanical models \cite{Wang2012,2013-TOG-MuscleBasedBipeds,Lee2014}. However, these models can be challenging to build, difficult to control, and may still result in unnatural behaviours. In light of these challenges, data-driven RL methods that utilize reference motion data have been proposed as an alternative 
\cite{Won2017,2018-TOG-DeepMimic}. Reference motion clips can be incorporated via a motion imitation objective that incentivizes the policy to produce behaviours that resemble the reference motions. In this paper, we explore methods for extending motion imitation with RL to accommodate low-fidelity reference motions extracted from videos, and introduce a novel adaptive state initialization technique that makes this practical even for highly dynamic and acrobatic movements.

\paragraph{Monocular Human Pose Estimation:}
While mocap remains the most popular source of demonstrations, it typically requires significant instrumentation, which limits its accessibility. Practitioners therefore often turn to public databases to satisfy their mocap needs \cite{CMUMocap,SFUMocap}. Unfortunately, the volume of publicly available mocap data is severely limited compared to datasets in other disciplines, such as ImageNet \cite{imagenet_cvpr09}. 
Alternatively, video clips are an abundant and accessible source of motion data. While recovering motion from raw video has been a long standing challenge \cite{Lee1985,bregler1998tracking}, recently deep learning approaches have made rapid progress in this area. 

Performance of 2D pose estimation improved rapidly after \citet{Toshev:2014} introduced a deep learning approach for predicting the 2D coordinates of joints directly from images. This is followed by methods that predict joint locations as a spatial heat map \cite{tompson2014joint,Wei:CVPR:2016,hourglass}.
In this work we build upon the recent OpenPose framework \cite{cao2017realtime}, which extends previous methods for real-time multi-person 2D pose estimation.
Monocular 3D pose estimation is an even more challenging problem due to depth ambiguity, which traditional methods resolve with strong priors \cite{Taylor:2000,Zhou:2015b,SMPLify}. 
The introduction of large-scale mocap datasets \cite{Human36m:2014} with ground truth 3D joint locations allowed for the development of deep learning based methods that directly estimate 3D joint locations from images \cite{Zhou:2016a,VNect,Pavlakos}. However, mocap datasets are typically captured in heavily instrumented environments, and models trained on these datasets alone do not generalize well to the complexity of images of humans \emph{in the wild}.
Therefore, recent methods focus on weakly supervised techniques, where a model may also be trained on images without ground truth 3D pose \cite{RogezMocap,Xingyi2017}. 
Note that most approaches only estimate the 3D joint \emph{locations} and not the 3D rotations of a kinematic tree, which is necessary to serve as reference for our RL algorithm.
Methods that predict joint locations require additional post-processing to recover the joint rotations through inverse kinematics \cite{VNect}. Only a handful of techniques directly estimate the 3D human pose as 3D joint rotations 
\cite{Xingyi2016,tung2017self,hmrKanazawa17}. 
Although there are methods that utilize video sequences as input \cite{Tekin:2015}, most state-of-the-art approaches predict the pose independently for each video frame. 
Recently \citet{MonoPerfCap_SIGGRAPH2018} propose a method that recovers a temporally consistent trajectory from monocular video by an additional optimization step in the 3D pose space. However, their method requires a pre-acquired template mesh of the actor and hence cannot be applied to legacy videos, such as those available from YouTube.
In this work we build on the recent work of \citet{hmrKanazawa17}, which is a weakly-supervised deep learning framework that trains a model to directly predict the 3D pose, as joint rotations, from a single image. 
A more detailed discussion is available in Section \ref{sec:pose}.
\paragraph{Video Imitation:} The problem of learning controllers from monocular video has received modest attention from the computer graphics community. The work most related to ours is the previous effort by \citet{Vondrak2012}, which demonstrated learning bipedal controllers for walking, jumping, and handsprings from videos. The controllers were represented as a finite-state machines (FSM), where the structure of the FSM and the parameters at each state were learned through an incremental optimization process. Manually-crafted balance strategies and inverse-dynamics models were incorporated into the control structure within each state of the FSM. To imitate the motion of the actor in a video, the controllers were trained by optimizing a 2D silhouette likelihood computed between the actor and simulated character. To resolve depth ambiguity, they incorporated a task-specific pose prior computed from mocap data. While the system was able to synthesize controllers for a number of skills from video demonstrations, the resulting motions can appear robotic and the use of a silhouette likelihood can neglect a significant amount of task-relevant information in the video. Furthermore, the task-pose priors require access to mocap clips that are similar to the skills being learned. If such data is already available, it might be advantageous to imitate the mocap clips instead. Similarly, \citet{2011-TOG-quadruped} utilized video clips of canine motions to train quadruped controllers, where the reference motions were extracted via manually annotating gait graphs and marker locations.

In this work, we take advantage of state-of-the-art 3D pose estimation techniques to extract full-body 3D reference motions from video, which resolves much of the depth ambiguity inherent in monocular images and improves the motion quality of the learned controllers. Deep RL enables the use of simple but general control structures that can be applied to a substantially wider range of skills, including locomotion, acrobatics, martial arts, and dancing. Our approach can be further extended to a novel physics-based motion completion application, where plausible future motions of a human actor can be predicted from a single image by leveraging a library of learned controllers. 
While our framework combines several components proposed in prior work, including the use of vision-based pose estimators \cite{Wei:CVPR:2016,hmrKanazawa17} and deep reinforcement learning with reference motion data~\cite{2018-TOG-DeepMimic}, the particular combination of these components is novel, and we introduce a number of extensions that are critical for integrating these disparate systems. To the best of our knowledge, the only prior work that has demonstrated learning full-body controllers from monocular video is the work by \citet{Vondrak2012}. Although incorporating reinforcement learning to imitate video demonstrations is conceptually natural, in practice it presents a number of challenges arising from nonphysical behaviours and other artifacts due to inaccurate pose estimation.

\section{Overview}

\begin{figure}[t]
	\centering
    \includegraphics[width=1\columnwidth]{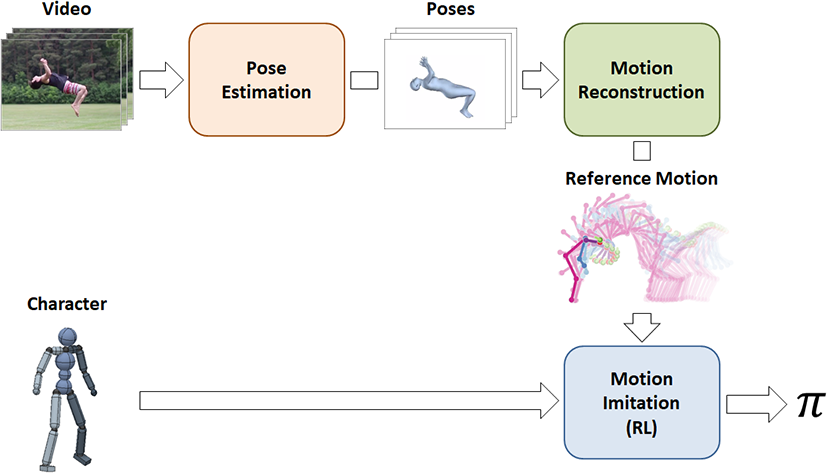}\\
\vspace{-0.25cm}
\caption{The pipeline consists of three stages: pose estimation, motion reconstruction, and imitation. It receives as input, a video clip of an actor performing a particular skill and a simulated character model, and outputs a control policy that enables the character to reproduce the skill in simulation.}
\label{fig:overview}
\end{figure}

Our framework receives as input a video clip and a simulated character model. It then synthesizes a controller that enables a physically simulated character to perform the skill demonstrated by the actor in the video. The resulting policies are robust to significant perturbations, can be retargeted to different characters and environments, and are usable in interactive settings.
The learning process is divided into three stages: pose estimation, motion reconstruction, and motion imitation. A schematic illustration of the framework is available in Figure~\ref{fig:overview}. The input video is first processed by the pose estimation stage, where a learned 2D and 3D pose estimators are applied to extract the pose of the actor in each frame. Next, the set of predicted poses proceeds to the motion reconstruction stage, where a reference motion trajectory $\{q^*_t\}$ is optimized such that it is consistent with both the 2D and 3D pose predictions, while also enforcing temporal-consistency between frames and mitigating other artifacts present in the original set of predicted poses.
The reference motion is then utilized in the motion imitation stage, where a control policy $\pi$ is trained to enable the character to reproduce the reference motion in a physically simulated environment. The pose estimator is trained with a weakly-supervised learning approach, and the control policy is trained with reinforcement learning using a motion imitation objective.

\section{Background}
\paragraph{Pose estimation:}
\label{sec:pose}
Our approach builds upon the recent 2D and 3D pose estimators, OpenPose
\cite{Wei:CVPR:2016} and Human Mesh Recovery (HMR) \cite{hmrKanazawa17} respectively. 
OpenPose performs both detection and 2D pose estimation of humans from a single image. It outputs the 2D pose as joint locations $x_j \in \mathbb{R}^{2}$ in the image coordinate space, as well as a confidence score for each joint $c_j \in \mathbb{R}$.
HMR is a recent approach that directly predicts the 3D pose and shape of a human model \cite{SMPL}, along with the camera configuration from an image of a localized person. 
The predicted 3D pose $q = \{q_j\}$ is parameterized by the local rotation of each joint $q_j$, represented in axis-angle form with respect to the parent link's coordinate frame. The world transformation of the root, designated to be the pelvis, is obtained using the predicted weak-perspective camera $\Pi$. 
The 3D pose is predicted 
by first encoding an image $I$ into a $2048D$ latent space $z = f(I)$ via a learned encoder $f$. The latent features are then decoded by a learned decoder $q(z)$ to produce the pose.
HMR uses a weakly-supervised adversarial framework that allows the model to be trained on images with only 2D pose annotations, \emph{without} any ground truth 3D labels. 
Therefore, it can be trained on datasets of in-the-wild images, such as COCO \cite{COCO}, and sports datasets \cite{LSP}, which is vital for learning acrobatic skills from video clips.

\paragraph{Reinforcement Learning:}
Our algorithm makes use of reinforcement learning, which has previously been used for imitation of mocap data \cite{2016-TOG-controlGraphs,2018-TOG-DeepMimic}.
During the motion imitation stage, the control policy is trained to imitate a reference motion via a motion imitation objective.
Training proceeds by having an agent interact with its environment according to a policy $\pi(a|s)$, which models the conditional distribution of action $a \in A$ given a state $s \in S$. At each timestep $t$, the agent observes the current state $s_t$ and samples an action $a_t$ from $\pi$. The environment then responds with a successor state $s' = s_{t+1}$, sampled from the dynamics $p(s' | s, a)$, and a scalar reward $r_t$, which reflects the desirability of the transition. For a parametric policy $\pi_\theta(a | s)$, with parameters $\theta$, the goal of the agent is to learn the optimal parameters $\theta^*$ that maximizes its expected return
\[J(\theta) = \mathbb{E}_{\tau \sim p_\theta(\tau)} \left[\sum_{t = 0}^T \gamma^t r_t \right],\]
where $p_\theta(\tau) = p(s_0) \prod_{t = 0}^{T - 1} p(s_{t + 1} | s_t, a_t) \pi_\theta(a_t | s_t)$ is the distribution over trajectories $\tau = (s_0, a_0, s_1, ..., a_{T - 1}, s_T)$ induced by the policy $\pi_\theta$, with $p(s_0)$ being the initial state distribution. $\sum_{t = 0}^T \gamma^t r_t$ represents the discounted return of a trajectory, with a horizon of $T$ steps and a discount factor $\gamma \in [0, 1]$.

Policy gradient methods are a popular class of algorithms for optimizing parametric policies \cite{sutton2001policy}. The algorithm optimizes $J(\theta)$ via gradient ascent, where the gradient of the objective with respect to the policy parameters $\triangledown_\theta J(\theta)$ is estimated using trajectories that are obtained by rolling out the policy:
\[\triangledown_\theta J(\theta) = \mathbb{E}_{s_t \sim d_\theta(s_t), a_t \sim \pi_\theta(a_t | s_t)} \left[ \triangledown_\theta \mathrm{log}(\pi_\theta(a_t | s_t)) \mathcal{A}_t \right], \]
where $d_\theta(s_t)$ is the state distribution under the policy $\pi_\theta$. $\mathcal{A}_t = R_t - V(s_t)$ represents the advantage of taking an action $a_t$ at a given state $s_t$, with $R_t = \sum_{l = 0}^{T - t} \gamma^l r_{t + l}$ being the return received by a particular trajectory starting from state $s_t$ at time $t$ and $V(s_t)$ is a value function that estimates the average return of starting in $s_t$ and following the policy for all subsequent steps. A number of practical improvements have been proposed, such as trust regions \cite{SchulmanLMJA15}, natural gradient \cite{Kakade2001}, and entropy regularization to prevent premature distribution collapse \cite{WilliamsPeng1991}.

\section{Pose Estimation}
Given a video clip, the role of the pose estimation stage is to predict the
pose of the actor in each frame. Towards this goal, there are two main challenges for our task. First, the acrobatic skills that we wish to imitate exhibit challenging poses that vary significantly from the distribution of common poses available in most datasets.
Second, poses are predicted independently for each frame, and therefore may not be temporally consistent, especially for dynamic motions. We address these challenges by leveraging an ensemble of pose estimators and a simple but effective data augmentation technique that substantially improve the quality of the predictions.

\begin{figure}[t!]
	\centering
    \includegraphics[width=1\columnwidth]{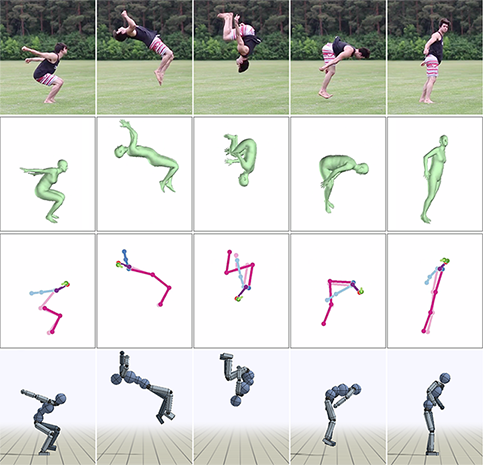}
     \vspace{-0.5cm}
\caption{Comparison of the motions generated by different stages of the pipeline for backflip A. \textbf{Top-to-Bottom:} Input video clip, 3D pose estimator, 2D pose estimator, simulated character.}
\label{fig:snapshotBackflip}
\vspace{-0.4cm}
\end{figure}

One of the challenges of tracking acrobatic movements is that they tend to exhibit complex poses with wildly varying body orientations (e.g.\ flips and spins). These poses are typically underrepresented in existing datasets, which are dominated by everyday images of humans in upright orientations. Thus, off-the-shelf pose estimators struggle to predict the poses in these videos.
To compensate for this discrepancy, we augment the standard datasets with rotated versions of the existing images, where the rotations are sampled uniformly between $[0, 2 \pi]$. We found that training the pose estimators with this augmented dataset, substantially improves performance for acrobatic poses.
Once trained, both estimators are applied independently to every frame to extract a 2D pose trajectory $\{\hat{x}_t\}$ and 3D pose trajectory $\{\hat{q}_t\}$. Note that the 2D pose $\hat{x}_t$ consists only of the 2D screen coordinates of the actor's joints, but tends to be more accurate than the 3D predictions. Examples of the predictions from the different pose estimators are shown in Figure~\ref{fig:snapshotBackflip}. The independent predictions from the pose estimators are then consolidated in the motion reconstruction stage to produce the final reference motion.

\section{Motion Reconstruction}
Since poses are predicted independently for every frame in the pose estimation stage, simply sequencing the poses into a trajectory tends to produce motions that exhibit artifacts due to inconsistent predictions across adjacent frames (see supplementary video).
The motion artifacts often manifest as nonphysical behaviours in the reference motion, such as high-frequency jitter and sudden changes in pose. These artifacts can hinder the simulated character's ability to reproduce the intended motion. The role of the motion reconstruction stage is to take advantage of the predictions from the two pose estimators to reconstruct a new kinematic trajectory that reconciles the individual predictions and mitigates artifacts, such that the resulting reference motion is more amenable for imitation. 

Specifically, given the predictions from the 2D and 3D pose estimators, we optimize a 3D pose trajectory that consolidates their predictions 
while also enforcing temporal consistency between adjacent frames.
Instead of directly optimizing in the 3D pose space, we take advantage of the encoder-decoder structure of the 3D pose estimator and optimize the 3D pose trajectory in the latent pose space $z_t$, which captures the manifold of 3D human poses \cite{hmrKanazawa17}.
The final 3D reference motion is constructed by optimizing a trajectory $Z = \{z_t\}$ in the latent space to minimize the reconstruction loss $l_{rec}$:
\[l_{rec}(Z) = w_{2D}l_{2D}(Z) + w_{3D}l_{3D}(Z) + w_{sm}l_{sm}(Z)\]
\[w_{2D} = 10, \\ w_{3D} = 100, \\ w_{sm} = 25,\]
The 2D consistency loss $l_{2D}$ minimizes the reprojection error between the predicted 2D joint locations and the 2D projections of the corresponding joints arising from the pose specified by $z_t$
\[l_{2D} = \sum_t \sum_j c_{t, j} \left\Vert \left(\hat{x}_{t,j} - \Pi \left[F_j\left(q\left(z_t \right)\right) \right] \right) \right\Vert_1,\] 
where $\hat{x}_{t,j}$ is the predicted 2D location of the $j$th joint, $c_{t,j}$ is the confidence of the prediction, and $F_j[\cdot]$ is the forward kinematics function that computes the 3D position of joint $j$ given the 3D pose. $q(z_{t})$ represents the pose decoded from $z_t$, and $\Pi\left[\cdot\right]$ is the weak-perspective projection that transforms 3D positions to 2D screen coordinates. 

The 3D consistency loss $l_{3D}$ encourages the optimized trajectory to stay close to the initial 3D prediction $\hat{q}_t$:
\[l_{3D} = \sum_t w_t dist(\hat{q}_{t}, q(z_t)),\]
where $dist(\cdot, \cdot)$ measures the distance between two rotations by the angle of the difference rotation. $w_t = \exp(-\delta_t)$ estimates the confidence of the initial 3D prediction using the difference between the initial 2D and 3D predictions, computed via the reprojection error  $\delta_t = \sum_j c_{t,j}||(\hat{x}_{t,j} - \Pi F_{j}(\hat{q}_t)) ||_2$. This ensures that initial 3D poses that are consistent with the 2D predictions are preserved, while inconsistent poses are adjusted through the other terms in the loss.

Finally, the smoothness loss $l_{sm}$ encourages smoothness of the 3D joint positions between adjacent frames
\[l_{sm} = \sum_{t} \sum_j \left\Vert F_j(q(z_t)) - F_j(q(z_{t+1})) \right\Vert_2^2. \]
After the optimization process, we obtain the final 3D reference motion $\{q^*_t\} = \{q(z^*_t)\}$. 



\section{Motion Imitation with RL}
Once the reference motion has been reconstructed, it progresses to the motion imitation stage, where the goal is to learn a policy $\pi$ that enables the character to reproduce the demonstrated skill in simulation.
The reference motion extracted by the previous stages is used to define an imitation objective, and a policy is then trained through reinforcement learning to imitate the given motion. The policy is modeled as a feedforward network that receives as input the current state $s$ and outputs an action distribution $\pi(a|s)$. 
To train the policy, we propose a variant of proximal policy optimization (PPO) \cite{PPO17} augmented with an adaptive initial state distribution, as described below.

\subsection{Initial State Distribution}

The initial state distribution $p(s_0)$ determines the states at which an agent starts each episode. Careful choice of $p(s_0)$ can have a significant impact on the performance of the resulting policy, as it can mitigate the challenges of exploration inherent in RL. An effective initial state distribution should expose the agent to promising states that are likely to maximize its expected return, thereby reducing the need for the agent to discover such states on its own. In the case of learning to imitate a reference motion, sampling initial states from the target trajectory can be highly effective for reproducing dynamic motions, such as flips and spins \cite{2018-TOG-DeepMimic}.
But the effectiveness of this strategy depends heavily on the quality of the reference motion. Due to artifacts from the pose estimator and modeling discrepancies between the simulated character and real-world actor, states sampled from the reference motion may not be ideal for reproducing the entirety of the motion.
For example, a common artifact present in motion data recorded from real-world actors, be it through motion capture or vision-based pose estimation, is high-frequency jittering, which can manifest as initial states with large joint velocities. Naively initializing the agent to such states will then require the agent to recover from the artifacts of the reference motion. These artifacts can be substantially more pronounced in reference motions reconstructed from video. Though the motion reconstruction stage is able to mitigate many of these artifacts, some errors may still persist.

While a myriad of post-processing techniques can be applied to mitigate artifacts in the reference motion, we can instead reduce the dependency on the quality of the reference motion by learning an initial state distribution with the specific purpose of aiding the character in learning an effective controller. This can be formulated as a cooperative multi-agent reinforcement learning problem where the first agent, defined by the policy $\pi_\theta(a_t | s_t)$, controls the movement of the character, and the second agent $\rho_\omega(s_0)$ proposes the initial states at which the character should start each episode. Both agents cooperate in order to maximize the multi-agent objective:
\[J(\theta, \omega) = \mathbb{E}_{\tau \sim p_{\theta, \omega}(\tau)} \left[\sum_{t = 0}^T \gamma^t r_t \right]\]
\[= \int_\tau \left( \rho_\omega(s_0) \prod_{t = 0}^{T - 1} p(s_{t + 1} | s_t, a_t) \pi_\theta(a_t | s_t) \right) \left(\sum_{t = 0}^T \gamma^t r_t \right) d\tau .\]
Note that, since the reward requires tracking the entire reference motion (as discussed in the next section), the initial state distribution cannot obtain the maximum return by ``cheating'' and providing excessively easy initializations. The maximum return is attained when the initial state distribution is close to the reference trajectory, but does not initialize the character in states from which recovery is impossible, as might be the case with erroneous states due to tracking error. The policy gradient of the initial state distribution $\rho_\omega(s_0)$ can be estimated according to:
\[ = \mathbb{E}_{\tau \sim p_{\theta, \omega}(\tau)} \left[\triangledown_\omega \mathrm{log} \left(\rho_\omega(s_0) \right) \sum_{t = 0}^T \gamma^t r_t \right]. \]
A detailed derivation is available in the supplementary material. Similar to the standard policy gradient for $\pi$, the gradient of the initial state distribution can be interpreted as increasing the likelihood of initial states that result in high returns. Unlike the standard policy gradient, which is calculated at every timestep, the gradient of the initial state distribution is calculated only at the first timestep. The discount factor captures the intuition that the effects of the initial state attenuates as the episode progresses. We will refer to this strategy of learning the initial state distribution as adaptive state initialization (ASI). Learning an initial state distribution can also be interpreted as a form of automatic curriculum generation, since $\rho_\omega(s_0)$ is incentivized to propose states that enable the character to closely reproduce the reference motion while also avoiding states that may be too challenging for the current policy.

\subsection{Reward}
The reward function is designed to encourage the character to match the reference motion $\{q^*_t\}$ generated by the motion reconstruction stage. The reward function is similar to the imitation objective proposed by \citet{2018-TOG-DeepMimic}, where at each step, the reward $r_t$ is calculated according to:
\[r_t = w^p r_t^p + w^v r_t^v + w^e r_t^e + w^c r_t^c\]
\[w^p = 0.65, w^v = 0.1, w^e = 0.15, w^c = 0.1.\]
The pose reward $r_t^p$ incentivizes the character to track the joint orientations from the reference motion, computed by quaternion differences of the simulated character's joint rotations and those of the reference motion. In the equation below, $q_{t,j}$ and $q^*_{t,j}$ represent the rotation of the $j$th joint from the simulated character and reference motion respectively, $q_1 \ominus q_2$ denotes the quaternion difference, and $||q||$ computes the scalar rotation of a quaternion about its axis:
\[r_t^p = \mathrm{exp} \left[-2 \left(\sum_j ||q^*_{t,j} \ominus q_{t,j}||^2\right) \right]. \]
Similarly, the velocity reward $r_t^v$ is calculated from the difference of local joint velocities, with $\dot{q}_{t,j}$ being the angular velocity of the $j$th joint. The target velocity $\dot{q}^*_{t,j}$ is computed from the reference motion via finite difference. 
\[r_t^v = \mathrm{exp} \left[-0.1 \left(\sum_j ||\dot{q}^*_{t,j} - \dot{q}_{t,j}||^2\right) \right]. \]
The end-effector reward $r_t^e$ encourages the character's hands and feet to match the positions specified by the reference motion. Here, $p_{t,e}$ denotes the 3D position with respect to the root of end-effector $e \in [$left foot, right foot, left hand, right hand$]$:
\[r_t^e = \mathrm{exp} \left[-40 \left(\sum_e ||p^*_{t,e} - p_{t,e}||^2\right) \right]. \]
Finally, $r_t^c$ penalizes deviations in the character's center-of-mass $c_t$ from that of the reference motion $c^*_t$:
\[r_t^c = \mathrm{exp} \left[-10 \left(||c^*_t - c_t||^2\right) \right]. \]
The positional qualities $p^*_{t,e}$ and $c^*_t$ are computed from the reference pose $q^*_t$ via forward kinematics.

\subsection{Training}
Training proceeds episodically, where at the start of each episode, the character is initialized to a state $s_0$ sampled from the initial state distribution $\rho_\omega(s_0)$.
A rollout is then simulated by sampling actions from the policy $\pi_\theta(a_t | s_t)$ at every step. An episode is simulated to a fixed time horizon or until a termination criteria has been triggered. Early termination is triggered whenever the character falls, which is detected as any link in the character's torso coming into contact with the ground. Once terminated, the policy receives 0 reward for all remaining timesteps in the episode. This termination criteria is disabled for contact-rich motions, such as rolling. Once a batch of data has been collected, minibatches are sampled from the dataset and used to update the policy and value networks. TD($\lambda$) is used to compute target values for the value network $V_\psi(s)$ \cite{Sutton1998} and GAE($\lambda$) is used to compute advantages for policy updates \cite{SchulmanMLJA15}. Please refer to the supplementary material for a more detailed summary of the learning algorithm.

\section{Motion Completion}
Once a collection of policies has been trained for a corpus of different skills, we can leverage these policies along with a physics-based simulation to predict the future motions of actors in new scenarios.
Given a single still image of a human actor, the goal of motion completion is to predict the actor's motion in subsequent timesteps. We denote the library of skills as $\{(\{q^i_t\}, \pi^i)\}$, where $\pi^i$ is the policy trained to imitate reference motion $\{q^i_t\}$. To predict the actor's motion, the target image is first processed by the 3D pose estimator to extract the pose of the actor $\bar{q}$. The extracted pose is then compared to every frame of each reference motion to select the reference motion and corresponding frame $q^{i^*}_{t^*}$ that is most similar to $\bar{q}$,
\[q^{i^*}_{t^*} = \mathop{\mathrm{arg \ min}}_{q^i_t} ||\bar{q} \ominus q^i_t||.\]
Next, the simulated character is initialized to the state defined by the pose $\bar{q}$ and, since still images do not provide any velocity information, we initialize the character's velocity to that of the selected reference motion $\dot{q}^{i^*}_{t^*}$. The motion is then simulated by applying the policy $\pi^{i^*}$ corresponding to the selected motion.

\begin{figure}[t]
	\centering
    \includegraphics[width=1\columnwidth]{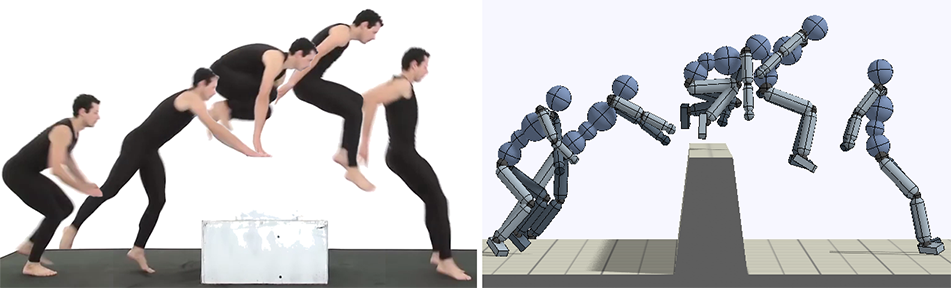}
    \vspace{-0.5cm}
\caption{Character imitating a 2-handed vault.}
\label{fig:snapshotVault}
\end{figure}

\section{Experimental Setup}
Our framework will be demonstrated using a 3D humanoid character and a simulated Atlas robot. The humanoid is modeled as articulated rigid bodies with a total of 12 joints, a mass of $45kg$, and a height of $1.62m$. Each link in the character's body is connected to its parent via a 3 degree-of-freedom spherical joint, except for the elbows and knees, which are modeled using a 1 degree-of-freedom revolute joint. The simulated Atlas robot follows a similar body structure, with a mass of $169.5kg$ and a height of of $1.82m$. The characters are actuated by PD controllers positioned at each joint, with manually specified gains and torque limits. 
Physics simulation is performed using the Bullet physics engine at $1.2kHz$ \cite{Bullet}, with the policy being queried at $30Hz$. All neural networks are built and trained using TensorFlow. Every episode is simulated for a maximum horizon of $20s$. Each policy is trained with approximately 150 million samples, taking about 1 day on a 16-core machine.

\begin{figure}[t]
	\centering
    \subfigure{\includegraphics[width=1\columnwidth]{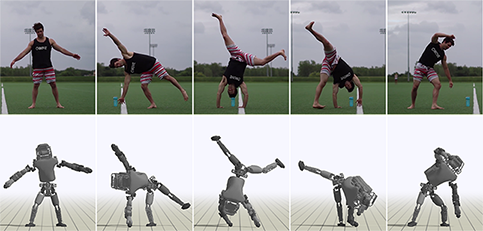}}\\
    \vspace{-0.25cm}
    \subfigure{\includegraphics[width=1\columnwidth]{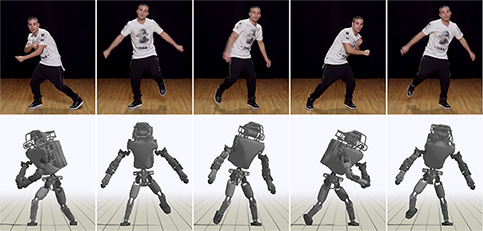}}\\
    \vspace{-0.25cm}
\caption{Simulated Atlas robot performing skills learned from video demonstrations. \textbf{Top:} Cartwheel A. \textbf{Bottom:} Dance.}
\label{fig:snapshotsAtlas}
\vspace{-0.25cm}
\end{figure}

\paragraph{Policy Network:}
The policy network $\pi(a|s)$ is constructed using 2 fully-connected layers with 1024 and 512 units respectively, with ReLU activations, followed by a linear output layer. 
The action distribution is modeled as a Gaussian with a state-dependent mean $\mu(s)$ and a fixed diagonal covariance matrix $\Sigma$:
\[\pi(a | s) = \mathcal{N}(\mu(s), \Sigma).\]
The value function $V_\psi(s)$ is represented by a similar network, with exception of the output layer, which produces a single scalar value.

\paragraph{State:} The state $s$ consists of features that describe the configuration of the character's body. The features are identical to those used by \citet{2018-TOG-DeepMimic}, which include the relative positions of each link with respect to the root, their rotations represented by quaternions, and their linear and angular velocities. All features are computed in the character's local coordinate frame, where the origin is located at the root and the x-axis pointing along the root link's facing direction. Since the target pose from the reference motion varies with time, a scalar phase variable $\phi \in [0, 1]$ is included among the state features. $\phi = 0$ denotes the start of the motion, and $\phi = 1$ denotes the end. The phase variables therefore helps to ensure that the policy is synchronized with the reference motion. Combined, the features constitute a $197D$ state space.

\paragraph{Action:} The action $a$ specifies target rotations for PD controllers positioned at each joint. 
For spherical joints, the targets are specified in a 4D axis-angle form. For revolute joints, the targets are specified by scalar rotation angles.
Combined, the parameters result in a $36D$ action space.

\begin{figure*}[t!]
	\centering
     \subfigure[Frontflip]{   \includegraphics[width=1.03\columnwidth]{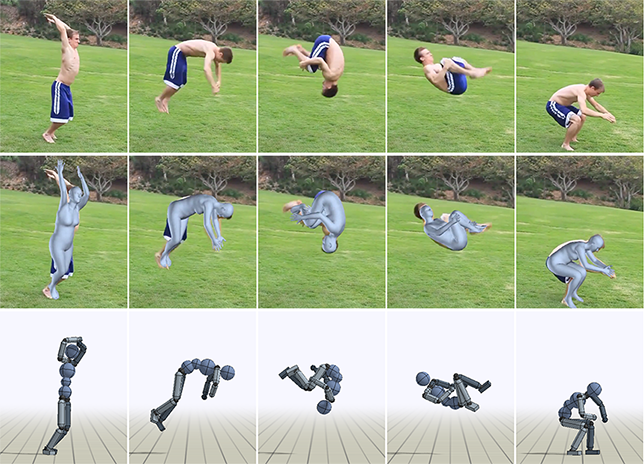}}
     \subfigure[Handspring A]{   \includegraphics[width=1.03\columnwidth]{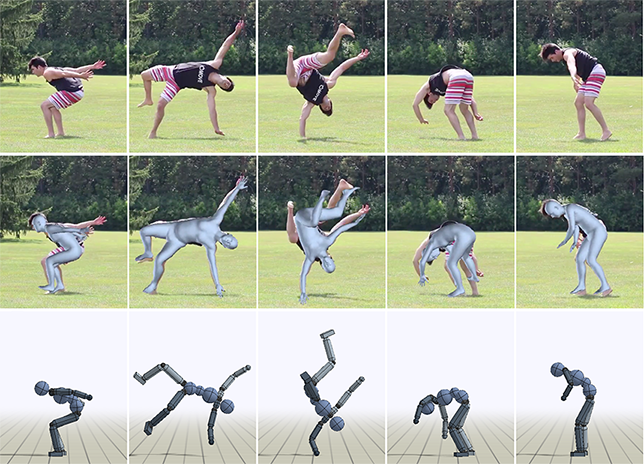}}\\
     \vspace{-0.25cm}
     \subfigure[Jump]{   \includegraphics[width=1.03\columnwidth]{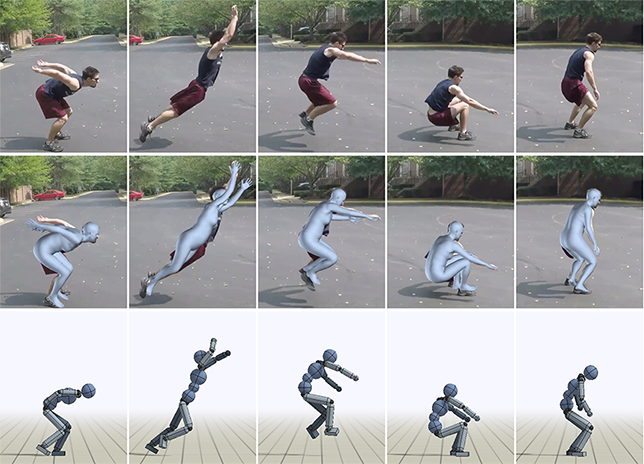}}
     \subfigure[Kip-Up]{   \includegraphics[width=1.03\columnwidth]{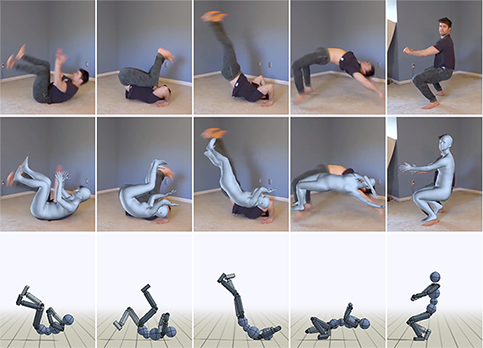}}\\
     \vspace{-0.25cm}
     \subfigure[Roll]{   \includegraphics[width=1.03\columnwidth]{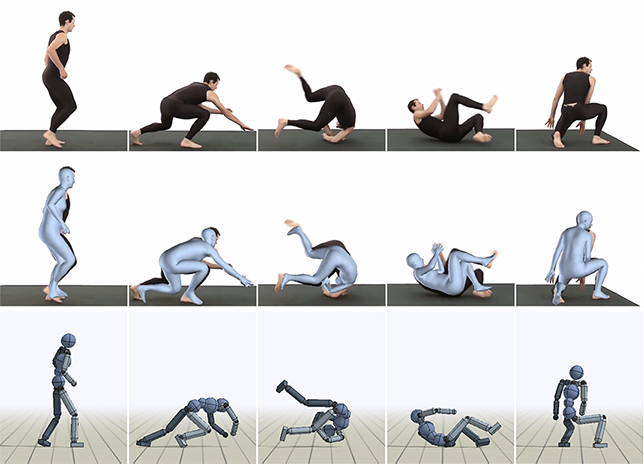}}
     \subfigure[Spin]{   \includegraphics[width=1.03\columnwidth]{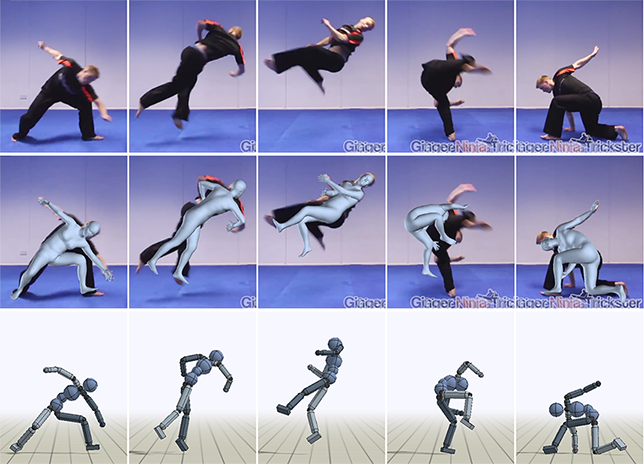}}\\
     \vspace{-0.5cm}
\caption{Simulated characters performing skills learned from video clips. \textbf{Top:} Video clip. \textbf{Middle:} 3D pose estimator. \textbf{Bottom:} Simulated character.}
\label{fig:snapshots0}
\end{figure*}

\paragraph{Initial State Distribution:}
At the start of each episode, the character is initialized to a state $s_0$ sampled from the initial state distribution $\rho_\omega(s_0)$. When applying adaptive state initialization, $\rho_\omega(s_0)$ is represented with a parametric model composed of independent Gaussian distributions over the character state. The Gaussian components are positioned at uniform points along the phase of the motion. To sample from this distribution, we first partition the state features $s = [\hat{s}, \phi]$, where $\phi$ is the phase variable and $\hat{s}$ represents the other features. The distribution $\rho_\omega(s)$ is then factorized according to:
\[\rho_\omega(s) = p_\omega(\hat{s} | \phi) p(\phi) ,\]
with $p(\phi)$ being a uniform distribution over discrete phase values $[\phi^0, \phi^1, ..., \phi^{k-1}]$. Each phase-conditioned state distribution $p_\omega(\hat{s} | \phi^i)$, corresponding to $\phi^i$, is modeled as a Gaussian $\mathcal{N}(\mu^i, \Sigma^i)$, with mean $\mu^i$ and diagonal covariance matrix $\Sigma^i$. The parameters of the initial state distribution consists of the parameters for each Gaussian component $\omega = \{\mu^i, \Sigma^i\}_{i=0}^{k-1}$. Both the mean and covariance matrix of each component are learned using policy gradients. When sampling an initial state, a phase value is first sampled from the discrete distribution $p(\phi)$. Next, $\hat{s}$ is sampled from $p_\omega(\hat{s} | \phi)$, and the combined features constitute the initial state.

\section{Results}

\begin{figure}[t]
	\centering
    \subfigure{   \includegraphics[width=1\columnwidth]{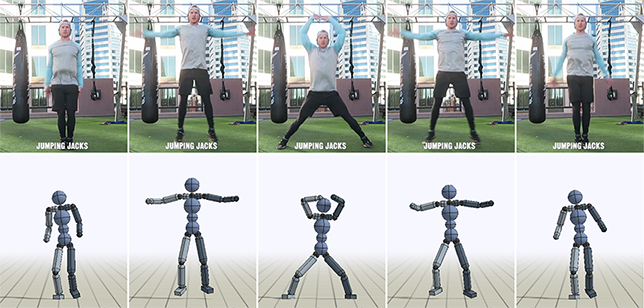}}\\
    \vspace{-0.25cm}
    \subfigure{   \includegraphics[width=1\columnwidth]{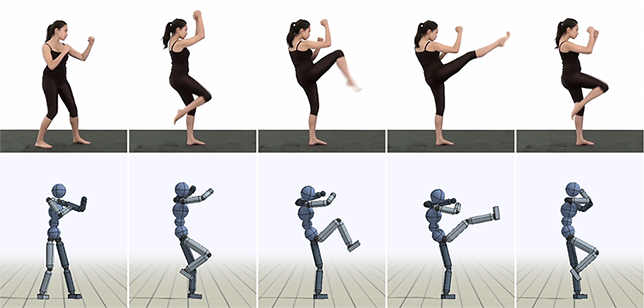}}\\
    \vspace{-0.25cm}
    \subfigure{   \includegraphics[width=1\columnwidth]{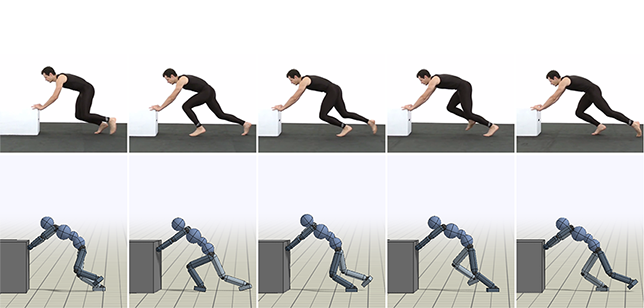}}\\
    \vspace{-0.25cm}
\caption{Humanoid character imitating skills from video demonstrations. \textbf{Top-to-Bottom:} Jumping jack, kick, push.}
\label{fig:snapshots1}
\vspace{-0.25cm}
\end{figure}

Motions from the trained policies are best seen in the \href{https://xbpeng.github.io/projects/SFV/index.html}{supplementary video}. Figures~\ref{fig:snapshotVault}, \ref{fig:snapshots0}, and \ref{fig:snapshots1} compare snapshots of the simulated characters with the original video demonstrations. All video clips were collected from YouTube. The clips depict human actors performing various acrobatic stunts (e.g. flips and cartwheels) and locomotion skills (walking and running). The clips were selected such that the camera is primarily stationary over the course of the motion, and only a single actor is present in the scene. 
Each clip is trimmed to contain only the relevant portions of their respective motion, and depicts one demonstration of a particular skill.

\begin{table}[t]
{ 
\centering  
\caption{Performance statistics of over 20 skills learned by our framework. $T_{cycle}$ denotes the length of the clip. $N_{samples}$ records the number of samples collected to train each policy. $NR$ represents the average normalized return of the final policy, with 0 and 1 being the minimum and maximum possible return per episode respectively.
For cyclic skills, the episode horizon is set to 20$s$. For acyclic skills, the horizon is determined by $T_{cycle}$. All statistics are recorded from the humanoid character unless stated otherwise.}
\vspace{-0.25cm}
\label{tab:perf}
\begin{tabular}{|l|c|c|c|}
\hline
{\bf Skill} & {\bf $\boldsymbol T_{cycle}$ (s)} & {\bf $\boldsymbol N_{samples}$} ($\boldsymbol 10^6$) & {\bf $\boldsymbol NR$} \\ \hline
Backflip A & 2.13 & 146 & 0.741 \\ \hline
Backflip B & 1.87 & 198 & 0.653 \\ \hline
Cartwheel A & 2.97 & 136 & 0.824 \\ \hline
Cartwheel B & 2.63 & 147 & 0.732 \\ \hline
Dance & 2.20 & 257 & 0.631 \\ \hline
Frontflip & 1.57 & 126 & 0.708 \\ \hline
Gangnam Style & 1.03 & 97 & 0.657 \\ \hline
Handspring A & 1.83 & 155 & 0.696 \\ \hline
Handspring B & 1.47 & 311 & 0.578 \\ \hline
Jump & 2.40 & 167 & 0.653 \\ \hline
Jumping Jack & 0.97 & 122 & 0.893 \\ \hline
Kick & 1.27 & 158 & 0.761 \\ \hline
Kip-Up & 1.87 & 123 & 0.788 \\ \hline
Punch & 1.17 & 115 & 0.831 \\ \hline
Push & 1.10 & 225 & 0.487 \\ \hline
Roll & 2.07 & 122 & 0.603 \\ \hline
Run & 0.73 & 126 & 0.878 \\ \hline
Spin & 1.07 & 146 & 0.779 \\ \hline
Spinkick & 1.87 & 196 & 0.747 \\ \hline
Vault & 1.43 & 107 & 0.730 \\ \hline
Walk & 0.87 & 122 & 0.932 \\ \hline
Atlas: Backflip A & 2.13 & 177 & 0.318 \\ \hline
Atlas: Cartwheel A & 2.97 & 174 & 0.456 \\ \hline
Atlas: Dance & 2.20 & 141 & 0.324 \\ \hline
Atlas: Handspring A & 1.83 & 115 & 0.360 \\ \hline
Atlas: Jump & 2.40 & 134 & 0.508 \\ \hline
Atlas: Run & 0.73 & 130 & 0.881 \\ \hline
Atlas: Vault & 1.43 & 112 & 0.752 \\ \hline
Atlas: Walk & 0.87 & 172 & 0.926 \\ \hline
\end{tabular} \\
}
\vspace{-0.25cm}
\end{table}

Table~\ref{tab:perf} summarizes the performance of the final policies, and a comprehensive set of learning curves are available in the supplementary material.
Performance is recorded as the average normalized return over multiple episodes. As it is challenging to directly quantify the difference between the motion of the actor in the video and the simulated character, performance is recorded with respect to the reconstructed reference motion. Since the reference motions recovered from the video clips may not be physically correct, a maximum return of 1 may not be achievable. Nonetheless, given a single video demonstration of each skill, the policies are able to reproduce a large variety of challenging skills ranging from contact-rich motions, such as rolling, to motions with significant flight phases, such as flips and spins. Policies can also be trained to perform skills that require more coordinated interactions with the environment, such as vaulting and pushing a large object.

\paragraph{Retargeting:}
One of the advantages of physics-based character animation is its ability to synthesize behaviours for novel situations that are not present in the original data. In addition to reproducing the various skills, our framework is also able to retarget the skills to characters and environments that differ substantially from what is presented in the video demonstrations. Since the same simulated character is trained to imitate motions from different human actors, the morphology of the character tends to differ drastically from that of the actor. To demonstrated the system's robustness to morphological discrepancies, we also trained a simulated Atlas robot to imitate a variety of video clips. The proportions of the Atlas' limbs differ significantly from normal human proportions, and with a weight of $169.5kg$, it is considerably heavier than the average human. Despite these drastic differences in morphology, our framework is able to learn policies that enable the Atlas to reproduce a diverse set of challenging skills. Table \ref{tab:perf} summarizes the performance of the Atlas policies, and Figure \ref{fig:snapshotsAtlas} illustrates snapshots of the simulated motions.

\begin{figure}[t]
	\centering
    \subfigure{   \includegraphics[width=1\columnwidth]{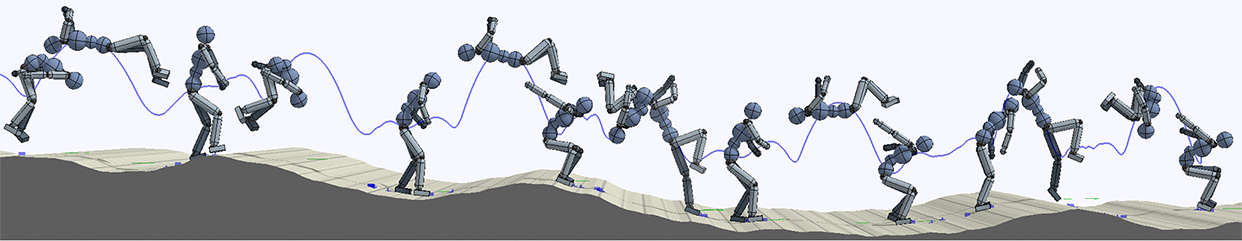}}\\
    \vspace{-0.25cm}
    \subfigure{   \includegraphics[width=1\columnwidth]{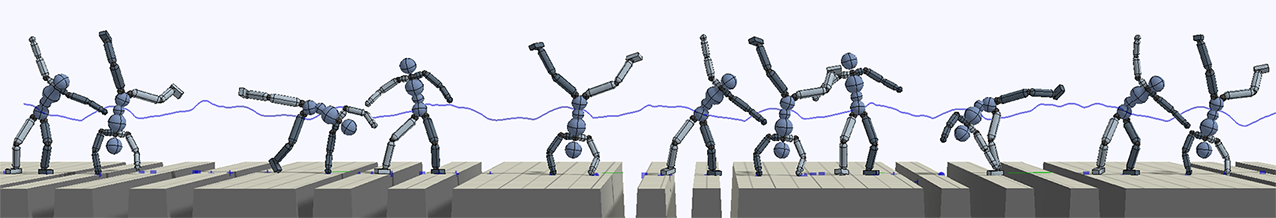}}\\
    \vspace{-0.25cm}
    \subfigure{   \includegraphics[width=1\columnwidth]{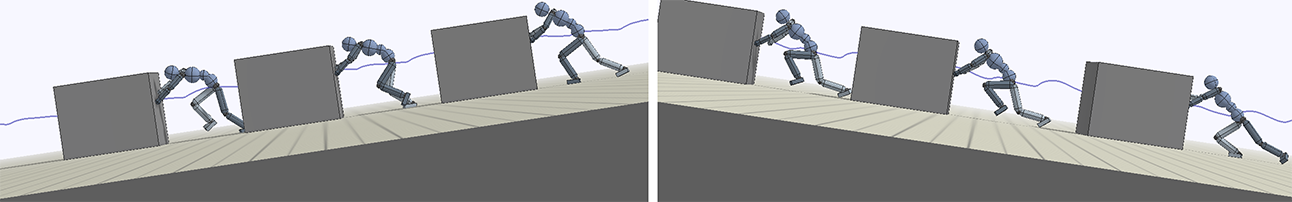}}
    \vspace{-0.25cm}
\caption{Skills retargeted to different environments. \textbf{Top-to-Bottom:} Backflip A across slopes, cartwheel B across gaps, pushing a box downhill and uphill.}
\label{fig:retargetTerrain}
\vspace{-0.25cm}
\end{figure}

In addition to retargeting to difference morphologies,
the skills can also be adapted to different environments. While the video demonstrations were recorded on flat terrain, our framework is able to train policies to perform the skills on irregular terrain. Figure~\ref{fig:retargetTerrain} highlights some of the skills that were adapted to environments composed of randomly generated slopes or gaps. The pushing skill can also be retargeted to push a $50kg$ box uphill and downhill with a slope of $15\%$. To enable the policies to perceive their environment, we follow the architecture used by \citet{2018-TOG-DeepMimic}, where a heightmap of the surrounding terrain is included in the input state, and the networks are augmented with corresponding convolutional layers to process the heightmap. Given a single demonstration of an actor performing a backflip on flat terrain, the policy is able to develop strategies for performing a backflip on randomly varying slopes. Similarly, the cartwheel policy learns to carefully coordinate the placement of the hands and feet to avoid falling into the gaps.

\begin{figure}[t]
	\centering
    \subfigure{   \includegraphics[width=0.48\columnwidth]{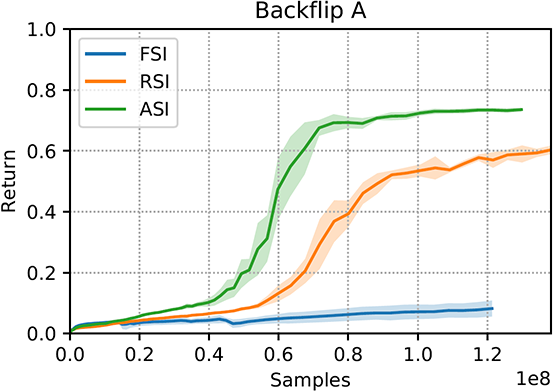}}
    \subfigure{   \includegraphics[width=0.48\columnwidth]{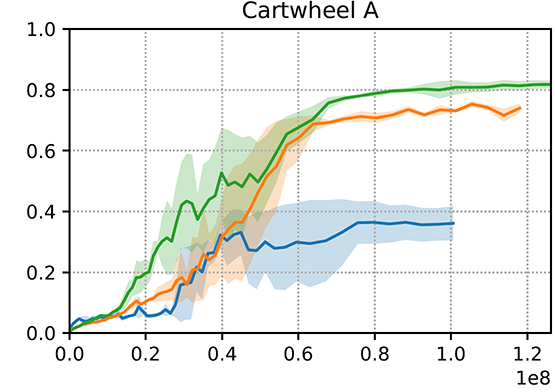}}\\
    \vspace{-0.15cm}
    \subfigure{   \includegraphics[width=0.48\columnwidth]{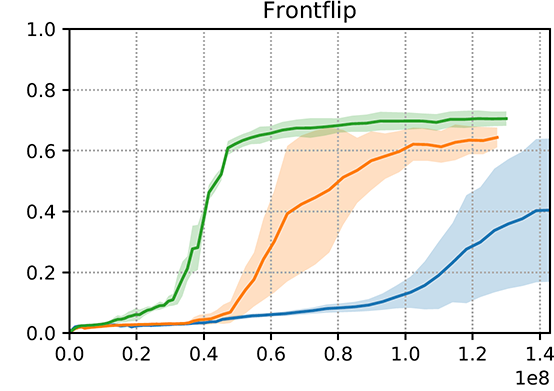}}
    \subfigure{   \includegraphics[width=0.48\columnwidth]{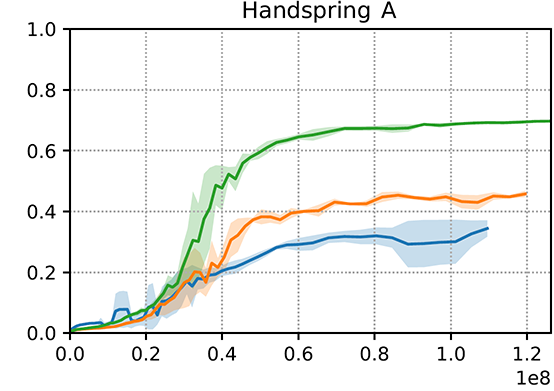}}
    \vspace{-0.5cm}
\caption{Learning curves comparing policies trained with fixed state initialization (FSI), reference state initialization (RSI), and adaptive state initialization (ASI). Three policies initialized with different random seeds are trained for each combination of skill and initial state distribution. Compared to its counterparts, ASI consistently improves performance and learning speed.}
\label{fig:curvesASI}
\end{figure}

\begin{table}[t]
{ 
\centering  
\caption{Performance of policies trained with different initial state distributions. ASI outperforms the other methods for all skills evaluated.}
\vspace{-0.25cm}
\label{tab:perfASI}
\begin{tabular}{|l|c|c|c|}
\hline
{\bf Skill} & {\bf FSI} & {\bf RSI} & {\bf ASI} \\ \hline
Backflip A & 0.086 & 0.602 & \textbf{0.741} \\ \hline
Cartwheel A & 0.362 & 0.738 & \textbf{0.824} \\ \hline
Frontflip & 0.435 & 0.658 & \textbf{0.708} \\ \hline
Handspring A & 0.358 & 0.464 & \textbf{0.696} \\ \hline
\end{tabular} \\
}
\vspace{-0.25cm}
\end{table}

\begin{figure*}[t!]
	\centering
    \subfigure[Backflip A]{\includegraphics[width=1.03\columnwidth]{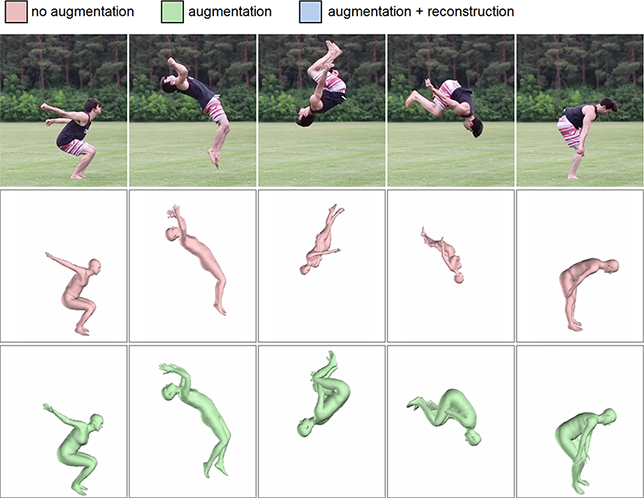}}
    \subfigure[Handspring A]{\includegraphics[width=1.03\columnwidth]{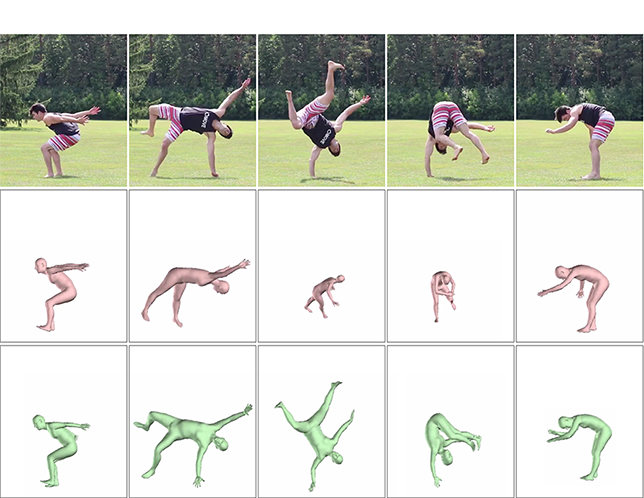}}\\
    \vspace{-0.25cm}
\caption{3D pose predictions with and without rotation augmentation. \textbf{Top:} Video. \textbf{Middle:} Without rotation augmentation. \textbf{Bottom:} With rotation augmentation. The pose estimator trained without rotation augmentation fails to correctly predict challenging poses, such as when the actor is upside-down.}
\label{fig:snapshotsRotAug}
\vspace{-0.2cm}
\end{figure*}

\begin{figure*}[t!]
	\centering
    \subfigure[Cartwheel A]{\includegraphics[width=1.03\columnwidth]{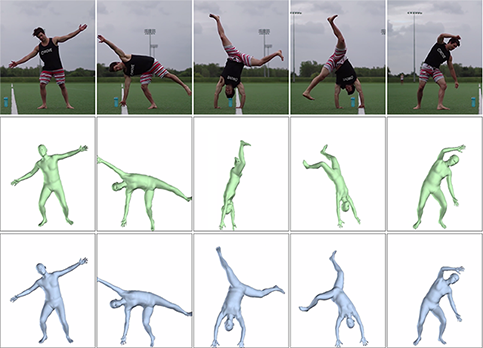}}
    \subfigure[Frontflip]{\includegraphics[width=1.03\columnwidth]{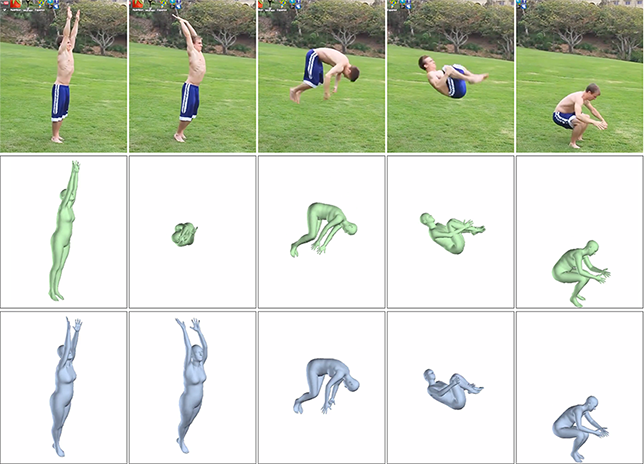}}\\
    \vspace{-0.25cm}
\caption{3D pose predictions before and after motion reconstruction. \textbf{Top:} Video. \textbf{Middle:} Raw predictions from the 3D pose estimator before motion reconstruction. \textbf{Bottom:} After motion reconstruction. The motion reconstruction process is able to fix erroneous predictions from the 3D pose estimator by taking advantage of the information from the 2D pose estimator and temporal consistency between adjacent frames.}
\label{fig:snapshotsMR}
\vspace{-0.2cm}
\end{figure*}

\paragraph{Initial State Distribution:}
To evaluate the impact of adaptive state initialization (ASI), we compare the performance of policies trained with ASI to those trained with fixed state initialization (FSI) and reference state initialization (RSI).
In the case of fixed state initialization, the character is always initialized to the same pose at the start of the motion. With reference state initialization, initial states are sampled randomly from the reference motion as proposed by \citet{2018-TOG-DeepMimic}. For ASI, the initial state distribution is modeled as a collection of $k=10$ independent Gaussian distributions positioned at uniformly spaced phase values. The mean of each Gaussian is initialized to the state at the corresponding phase of the reference motion, and the diagonal covariance matrix is initialized with the sample covariance of the states from the entire reference motion.
Both the mean and covariance matrix of each distribution are then learned through the training process, while the corresponding phase for each distribution is kept fixed. Figure~\ref{fig:curvesASI} compares the learning curves using the three different methods and Table~\ref{tab:perfASI} compares the performance of the final policies. Each result is averaged over three independent runs with different random seeds.

\begin{figure}[t]
	\centering
    \subfigure{   \includegraphics[width=0.48\columnwidth]{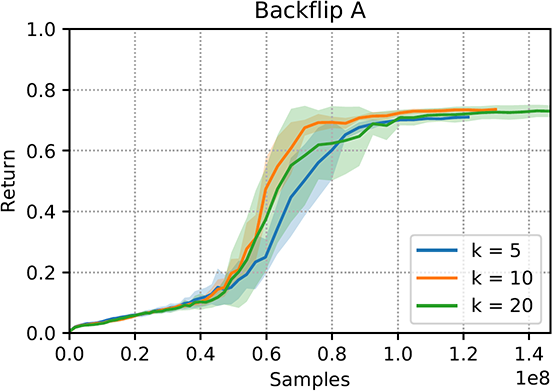}}
    \subfigure{   \includegraphics[width=0.48\columnwidth]{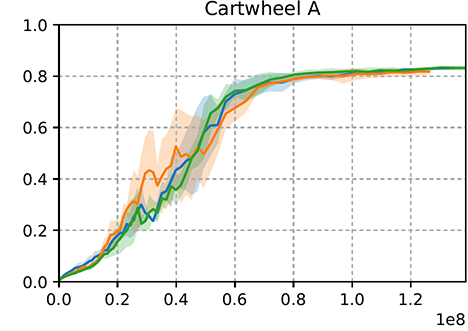}}
    \vspace{-0.5cm}
\caption{Learning curves of policies trained with ASI using different number of Gaussian components. The choice of the number of components does not appear to have a significant impact on performance.}
\label{fig:curvesASIcomp}
    \vspace{-0.3cm}
\end{figure}

Overall, the behaviour of the learning algorithm appears consistent across multiple runs. Policies trained with ASI consistently outperform their counterparts, converging to the highest return between the different methods. For more challenging skills, such as the backflip and frontflip, ASI also shows notable improvements in learning speed. Policies trained with FSI struggles to reproduce any of the skills. Furthermore, we evaluate the sensitivity of ASI to different numbers of Gaussian components. Policies were trained using $k={5, 10, 20}$ components and their corresponding learning curves are available in Figure~\ref{fig:curvesASIcomp}. Using different numbers of components does not seem to have a significant impact on the performance of ASI. Qualitatively, the resulting motions also appear similar. The experiments suggest that ASI is fairly robust to different choices for this hyperparameter.

\paragraph{Reference Motion:}
A policy's ability to reproduce a video demonstration relies on the quality of the reconstructed reference motion. Here, we investigate the effects of rotation augmentation and motion reconstruction on the resulting reference motions.
Most existing datasets of human poses are biased heavily towards upright poses. However, 
an actor's orientation can vary more drastically when performing highly dynamic and acrobatic skills, e.g.\ upside-down poses during a flip. Rotation augmentation significantly improves predictions for these less common poses.
Figure~\ref{fig:snapshotsRotAug} compares the predictions from pose estimators trained with and without rotation augmentation. We found that this step is vital for accurate predictions of more extreme poses, such as those present in the backflip and handspring. Without augmentation, both pose estimators consistently fail to predict upside-down poses.

\begin{figure}[t]
	\centering
    \subfigure{   \includegraphics[width=0.48\columnwidth]{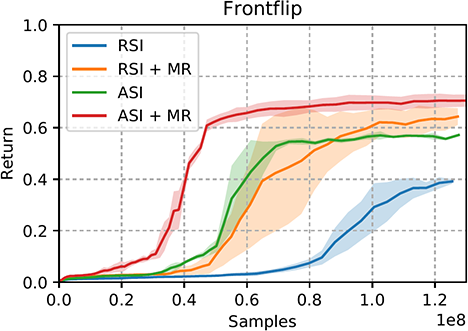}}
    \subfigure{   \includegraphics[width=0.48\columnwidth]{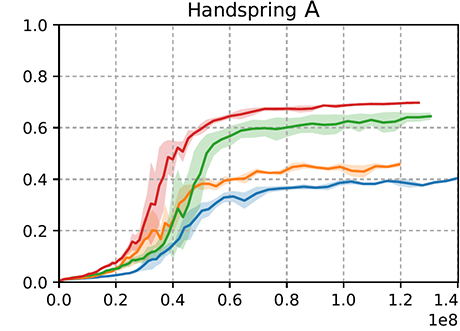}}\\
    \vspace{-0.5cm}
\caption{Learning curves comparing policies trained with and without motion reconstruction (MR). MR improves performance for both RSI and ASI.}
\label{fig:curvesMRcomp}
\end{figure}

\begin{table}[t]
{ 
\centering  
\caption{Performance of policies with and without motion reconstruction.}
\vspace{-0.15cm}
\label{tab:perfMR}
\begin{tabular}{|l|c|c|c|c|}
\hline
{\bf Skill} & {\bf RSI} & {\bf RSI + MR} & {\bf ASI} & {\bf ASI + MR} \\ \hline
Frontflip & 0.404 & 0.658 & 0.403 & \textbf{0.708} \\ \hline
Handspring A & 0.391 & 0.464 & 0.631 & \textbf{0.696} \\ \hline
\end{tabular} \\
}
\vspace{-0.25cm}
\end{table}

Next, we evaluate the effects of the motion reconstruction stage in producing reference motions that can be better reproduced by a simulated character. Polices that are trained to imitate the optimized reference motions generated by motion reconstruction (MR), are compared to policies trained without MR, where the poses from the 3D pose estimator are directly used as the reference motion. Figure~\ref{fig:snapshotsMR} compares the motions before and after MR. While the 3D pose estimator occasionally produces erroneous predictions, the MR process is able to correct these errors by taking advantage of the predictions from the 2D estimator and enforcing temporal consistency between adjacent frames. Learning curves comparing policies trained using reference motions before and after motion reconstruction are available in Figure~\ref{fig:curvesMRcomp}, and Table~\ref{tab:perfMR} summarizes the performance of the final policies. For each type of reference motion, we also compared policies trained with either RSI or ASI. Overall, imitating reference motions generated by the motion reconstruction processes improves performance and learning speed for the different skills. The improvements due to MR appears more pronounced when policies are trained with RSI. Since the initial states are sampled directly from the reference motion, performance is more susceptible to artifacts present in the reference motion. MR also shows consistent improvement across multiple training runs when using ASI. 
Note that since the reward reflects similarly to the reference motion, and not the original video, a higher return does not necessarily imply better reproduction of the video demonstration. Instead, the higher return with MR indicates that the simulated character is able to better reproduce the reference motions produced by MR than the raw predictions from the pose estimator. Thus, the results suggest that by enforcing temporal consistency and mitigating artifacts due to inaccurate pose predictions, motion reconstruction is able to generate reference motions that are more amenable to being mimicked by a simulated character.

\begin{figure}[t]
	\centering
    \subfigure{\includegraphics[width=1\columnwidth]{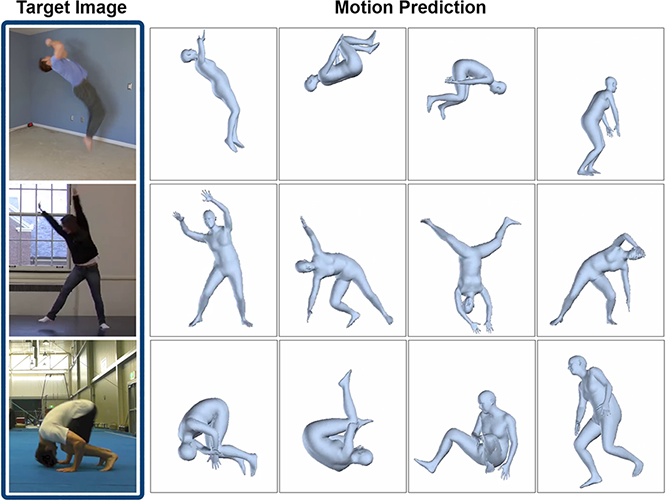}}\\
    \vspace{-0.4cm}
\caption{Given a target image, our motion completion technique predicts plausible motions for the actor in the image.}
\label{fig:motionCompletion}
\vspace{-0.5cm}
\end{figure}

\paragraph{Motion Completion:}
The motion completion technique is demonstrated on images depicting actors performing various acrobatic skills, such as backflips, cartwheels, and rolls. Figure~\ref{fig:motionCompletion} illustrates some of the predicted motions. Only a single image is provided during evaluation, and the system is then able to predict a plausible future motion for the depicted actor. The robustness of the learned policies enable the character to synthesize plausible behaviours even when the actor's pose $\bar{q}$ differs significantly from the reference motions used to train the policies. However, the system can fail to generate reasonable motions for an image if the actor's pose is drastically different from those of the reference motions, and the predictions are limited to skills spanned by the existing policies.

\section{Discussion and Limitations}
We presented a framework for learning full-body motion skills from monocular video demonstrations. Our method is able to reproduce a diverse set of highly dynamic and acrobatic skills with simulated humanoid characters. We proposed a data augmentation technique that improves the performance of pose estimators for challenging acrobatic motions, and a motion reconstruction method that leverages an ensemble of pose estimators to produce higher-fidelity reference motions. Our adaptive state initialization method substantially improves the performance of the motion imitation process when imitating low-fidelity reference motions. Our framework is also able to retarget skills to characters and environments that differ drastically from those present in the original video clips. By leveraging a library of learned controllers, we introduced a physics-based motion completion technique, where given a single image of a human actor, our system is able to predict plausible future motions of the actor.

While our framework is able to imitate a diverse collection of video clips, it does have a number of limitations. Since the success of the motion imitation stage depends on the accuracy of the reconstructed motion, when the pose estimators are not able to correctly predict an actor's pose, the resulting policy will fail to reproduce the behavior. Examples include the kip-up, where the reconstructed motions did not accurately capture the motion of the actor's arms, and the spinkick, where the pose estimator did not capture the extension of the actor's leg during the kick.
Furthermore, our characters still sometimes exhibit artifacts such as peculiar postures and stiff movements. Fast dance steps, such as those exhibited in the Gangnam Style clip, remains challenging for the system, and we have yet to be able to train policies that can closely reproduce such nimble motions. Due to difficulties in estimating the global translation of the character's root, our results have primarily been limited to video clips with minimal camera motion.

Nonetheless, we believe this work opens many exciting directions for future exploration. Our experiments suggest that learning highly-dynamic skills from video demonstrations is achievable by building on state-of-the-art techniques from computer vision and reinforcement learning.
An advantage of our modular design is that new advances relevant to the various stages of the pipeline can be readily incorporated to improve the overall effectiveness of the framework. However, an exciting direction for future work is to investigate methods for more end-to-end learning from visual demonstrations, for example taking inspiration from \citet{SermanetLHL17} and \citet{Yu2018}, which may reduce the dependence on accurate pose estimators. Another exciting direction is to capitalize on our method's ability to learn from video clips and focus on large, outdoor activities, as well as motions of nonhuman animals that are conventionally very difficult, if not impossible, to mocap. 

\begin{acks}
We thank the anonymous reviewers for their helpful feedback, and AWS for providing computational resources. This research was funded by an NSERC Postgraduate Scholarship, a Berkeley Fellowship for Graduate Study, and BAIR.
\end{acks}

\bibliographystyle{ACM-Reference-Format}
\bibliography{video_imitation}


\begin{thebibliography}{66}


\ifx \showCODEN    \undefined \def \showCODEN     #1{\unskip}     \fi
\ifx \showDOI      \undefined \def \showDOI       #1{#1}\fi
\ifx \showISBNx    \undefined \def \showISBNx     #1{\unskip}     \fi
\ifx \showISBNxiii \undefined \def \showISBNxiii  #1{\unskip}     \fi
\ifx \showISSN     \undefined \def \showISSN      #1{\unskip}     \fi
\ifx \showLCCN     \undefined \def \showLCCN      #1{\unskip}     \fi
\ifx \shownote     \undefined \def \shownote      #1{#1}          \fi
\ifx \showarticletitle \undefined \def \showarticletitle #1{#1}   \fi
\ifx \showURL      \undefined \def \showURL       {\relax}        \fi
\providecommand\bibfield[2]{#2}
\providecommand\bibinfo[2]{#2}
\providecommand\natexlab[1]{#1}
\providecommand\showeprint[2][]{arXiv:#2}

\bibitem[\protect\citeauthoryear{Aslam}{Aslam}{2018}]%
        {YouTubeStats}
\bibfield{author}{\bibinfo{person}{Salman Aslam}.}
  \bibinfo{year}{2018}\natexlab{}.
\newblock \bibinfo{title}{YouTube by the Numbers}.
\newblock
  \bibinfo{howpublished}{\url{https://www.omnicoreagency.com/youtube-statistics/}}.
    (\bibinfo{year}{2018}).
\newblock
\newblock
\shownote{Accessed: 2018-05-15.}


\bibitem[\protect\citeauthoryear{Bogo, Kanazawa, Lassner, Gehler, Romero, and
  Black}{Bogo et~al\mbox{.}}{2016}]%
        {SMPLify}
\bibfield{author}{\bibinfo{person}{Federica Bogo}, \bibinfo{person}{Angjoo
  Kanazawa}, \bibinfo{person}{Christoph Lassner}, \bibinfo{person}{Peter
  Gehler}, \bibinfo{person}{Javier Romero}, {and} \bibinfo{person}{Michael~J.
  Black}.} \bibinfo{year}{2016}\natexlab{}.
\newblock \showarticletitle{Keep it {SMPL}: Automatic Estimation of {3D} Human
  Pose and Shape from a Single Image}. In \bibinfo{booktitle}{\emph{European
  Conference on Computer Vision, ECCV}} \emph{(\bibinfo{series}{Lecture Notes
  in Computer Science})}. \bibinfo{publisher}{Springer International
  Publishing}.
\newblock


\bibitem[\protect\citeauthoryear{Bregler and Malik}{Bregler and Malik}{1998}]%
        {bregler1998tracking}
\bibfield{author}{\bibinfo{person}{Christoph Bregler} {and}
  \bibinfo{person}{Jitendra Malik}.} \bibinfo{year}{1998}\natexlab{}.
\newblock \showarticletitle{Tracking people with twists and exponential maps}.
  In \bibinfo{booktitle}{\emph{IEEE Conference on Computer Vision and Pattern
  Recognition, CVPR}}. IEEE, \bibinfo{pages}{8--15}.
\newblock


\bibitem[\protect\citeauthoryear{Brockman, Cheung, Pettersson, Schneider,
  Schulman, Tang, and Zaremba}{Brockman et~al\mbox{.}}{2016}]%
        {BrockmanCPSSTZ16}
\bibfield{author}{\bibinfo{person}{Greg Brockman}, \bibinfo{person}{Vicki
  Cheung}, \bibinfo{person}{Ludwig Pettersson}, \bibinfo{person}{Jonas
  Schneider}, \bibinfo{person}{John Schulman}, \bibinfo{person}{Jie Tang},
  {and} \bibinfo{person}{Wojciech Zaremba}.} \bibinfo{year}{2016}\natexlab{}.
\newblock \showarticletitle{OpenAI Gym}.
\newblock \bibinfo{journal}{\emph{CoRR}}  \bibinfo{volume}{abs/1606.01540}
  (\bibinfo{year}{2016}).
\newblock
\showeprint[arxiv]{1606.01540}


\bibitem[\protect\citeauthoryear{Bullet}{Bullet}{2015}]%
        {Bullet}
\bibfield{author}{\bibinfo{person}{Bullet}.} \bibinfo{year}{2015}\natexlab{}.
\newblock \bibinfo{title}{Bullet Physics Library}.
\newblock   (\bibinfo{year}{2015}).
\newblock
\newblock
\shownote{http://bulletphysics.org.}


\bibitem[\protect\citeauthoryear{Cao, Simon, Wei, and Sheikh}{Cao
  et~al\mbox{.}}{2017}]%
        {cao2017realtime}
\bibfield{author}{\bibinfo{person}{Zhe Cao}, \bibinfo{person}{Tomas Simon},
  \bibinfo{person}{Shih-En Wei}, {and} \bibinfo{person}{Yaser Sheikh}.}
  \bibinfo{year}{2017}\natexlab{}.
\newblock \showarticletitle{Realtime Multi-Person 2D Pose Estimation using Part
  Affinity Fields}. In \bibinfo{booktitle}{\emph{CVPR}}.
\newblock


\bibitem[\protect\citeauthoryear{CMU}{CMU}{2018}]%
        {CMUMocap}
\bibfield{author}{\bibinfo{person}{CMU}.} \bibinfo{year}{2018}\natexlab{}.
\newblock \bibinfo{title}{CMU Graphics Lab Motion Capture Database}.
\newblock   (\bibinfo{year}{2018}).
\newblock
\newblock
\shownote{http://mocap.cs.cmu.edu.}


\bibitem[\protect\citeauthoryear{Coros, Beaudoin, and van~de Panne}{Coros
  et~al\mbox{.}}{2009}]%
        {Coros09}
\bibfield{author}{\bibinfo{person}{Stelian Coros}, \bibinfo{person}{Philippe
  Beaudoin}, {and} \bibinfo{person}{Michiel van~de Panne}.}
  \bibinfo{year}{2009}\natexlab{}.
\newblock \showarticletitle{Robust Task-based Control Policies for
  Physics-based Characters}.
\newblock \bibinfo{journal}{\emph{ACM Trans. Graph. (Proc. SIGGRAPH Asia)}}
  \bibinfo{volume}{28}, \bibinfo{number}{5} (\bibinfo{year}{2009}),
  \bibinfo{pages}{Article 170}.
\newblock


\bibitem[\protect\citeauthoryear{Coros, Beaudoin, and van~de Panne}{Coros
  et~al\mbox{.}}{2010}]%
        {2010-TOG-gbwc}
\bibfield{author}{\bibinfo{person}{Stelian Coros}, \bibinfo{person}{Philippe
  Beaudoin}, {and} \bibinfo{person}{Michiel van~de Panne}.}
  \bibinfo{year}{2010}\natexlab{}.
\newblock \showarticletitle{Generalized Biped Walking Control}.
\newblock \bibinfo{journal}{\emph{ACM Transctions on Graphics}}
  \bibinfo{volume}{29}, \bibinfo{number}{4} (\bibinfo{year}{2010}),
  \bibinfo{pages}{Article 130}.
\newblock


\bibitem[\protect\citeauthoryear{Coros, Karpathy, Jones, Reveret, and van~de
  Panne}{Coros et~al\mbox{.}}{2011}]%
        {2011-TOG-quadruped}
\bibfield{author}{\bibinfo{person}{Stelian Coros}, \bibinfo{person}{Andrej
  Karpathy}, \bibinfo{person}{Ben Jones}, \bibinfo{person}{Lionel Reveret},
  {and} \bibinfo{person}{Michiel van~de Panne}.}
  \bibinfo{year}{2011}\natexlab{}.
\newblock \showarticletitle{Locomotion Skills for Simulated Quadrupeds}.
\newblock \bibinfo{journal}{\emph{ACM Transactions on Graphics}}
  \bibinfo{volume}{30}, \bibinfo{number}{4} (\bibinfo{year}{2011}),
  \bibinfo{pages}{Article TBD}.
\newblock


\bibitem[\protect\citeauthoryear{da~Silva, Abe, and Popovi\'{c}}{da~Silva
  et~al\mbox{.}}{2008}]%
        {daSilva2008a}
\bibfield{author}{\bibinfo{person}{Marco da Silva}, \bibinfo{person}{Yeuhi
  Abe}, {and} \bibinfo{person}{Jovan Popovi\'{c}}.}
  \bibinfo{year}{2008}\natexlab{}.
\newblock \showarticletitle{Interactive Simulation of Stylized Human
  Locomotion}. In \bibinfo{booktitle}{\emph{ACM SIGGRAPH 2008 Papers}}
  \emph{(\bibinfo{series}{SIGGRAPH '08})}. \bibinfo{publisher}{ACM},
  \bibinfo{address}{New York, NY, USA}, Article \bibinfo{articleno}{82},
  \bibinfo{numpages}{10}~pages.
\newblock
\showISBNx{978-1-4503-0112-1}
\urldef\tempurl%
\url{https://doi.org/10.1145/1399504.1360681}
\showDOI{\tempurl}


\bibitem[\protect\citeauthoryear{Deng, Dong, Socher, Li, Li, and Fei-Fei}{Deng
  et~al\mbox{.}}{2009}]%
        {imagenet_cvpr09}
\bibfield{author}{\bibinfo{person}{J. Deng}, \bibinfo{person}{W. Dong},
  \bibinfo{person}{R. Socher}, \bibinfo{person}{L.-J. Li}, \bibinfo{person}{K.
  Li}, {and} \bibinfo{person}{L. Fei-Fei}.} \bibinfo{year}{2009}\natexlab{}.
\newblock \showarticletitle{{ImageNet: A Large-Scale Hierarchical Image
  Database}}. In \bibinfo{booktitle}{\emph{CVPR09}}.
\newblock


\bibitem[\protect\citeauthoryear{Duan, Chen, Houthooft, Schulman, and
  Abbeel}{Duan et~al\mbox{.}}{2016}]%
        {DuanCHSA16}
\bibfield{author}{\bibinfo{person}{Yan Duan}, \bibinfo{person}{Xi Chen},
  \bibinfo{person}{Rein Houthooft}, \bibinfo{person}{John Schulman}, {and}
  \bibinfo{person}{Pieter Abbeel}.} \bibinfo{year}{2016}\natexlab{}.
\newblock \showarticletitle{Benchmarking Deep Reinforcement Learning for
  Continuous Control}.
\newblock \bibinfo{journal}{\emph{CoRR}}  \bibinfo{volume}{abs/1604.06778}
  (\bibinfo{year}{2016}).
\newblock
\showeprint[arxiv]{1604.06778}


\bibitem[\protect\citeauthoryear{Geijtenbeek, van~de Panne, and van~der
  Stappen}{Geijtenbeek et~al\mbox{.}}{2013}]%
        {2013-TOG-MuscleBasedBipeds}
\bibfield{author}{\bibinfo{person}{Thomas Geijtenbeek},
  \bibinfo{person}{Michiel van~de Panne}, {and} \bibinfo{person}{A.~Frank
  van~der Stappen}.} \bibinfo{year}{2013}\natexlab{}.
\newblock \showarticletitle{Flexible Muscle-Based Locomotion for Bipedal
  Creatures}.
\newblock \bibinfo{journal}{\emph{ACM Transactions on Graphics}}
  \bibinfo{volume}{32}, \bibinfo{number}{6} (\bibinfo{year}{2013}).
\newblock


\bibitem[\protect\citeauthoryear{Holden, Komura, and Saito}{Holden
  et~al\mbox{.}}{2017}]%
        {Holden2017}
\bibfield{author}{\bibinfo{person}{Daniel Holden}, \bibinfo{person}{Taku
  Komura}, {and} \bibinfo{person}{Jun Saito}.} \bibinfo{year}{2017}\natexlab{}.
\newblock \showarticletitle{Phase-functioned Neural Networks for Character
  Control}.
\newblock \bibinfo{journal}{\emph{ACM Trans. Graph.}} \bibinfo{volume}{36},
  \bibinfo{number}{4}, Article \bibinfo{articleno}{42} (\bibinfo{date}{July}
  \bibinfo{year}{2017}), \bibinfo{numpages}{13}~pages.
\newblock
\showISSN{0730-0301}


\bibitem[\protect\citeauthoryear{Holden, Saito, and Komura}{Holden
  et~al\mbox{.}}{2016}]%
        {Holden2016}
\bibfield{author}{\bibinfo{person}{Daniel Holden}, \bibinfo{person}{Jun Saito},
  {and} \bibinfo{person}{Taku Komura}.} \bibinfo{year}{2016}\natexlab{}.
\newblock \showarticletitle{A Deep Learning Framework for Character Motion
  Synthesis and Editing}.
\newblock \bibinfo{journal}{\emph{ACM Trans. Graph.}} \bibinfo{volume}{35},
  \bibinfo{number}{4}, Article \bibinfo{articleno}{138} (\bibinfo{date}{July}
  \bibinfo{year}{2016}), \bibinfo{numpages}{11}~pages.
\newblock
\showISSN{0730-0301}


\bibitem[\protect\citeauthoryear{Ionescu, Papava, Olaru, and
  Sminchisescu}{Ionescu et~al\mbox{.}}{2014}]%
        {Human36m:2014}
\bibfield{author}{\bibinfo{person}{Catalin Ionescu}, \bibinfo{person}{Dragos
  Papava}, \bibinfo{person}{Vlad Olaru}, {and} \bibinfo{person}{Cristian
  Sminchisescu}.} \bibinfo{year}{2014}\natexlab{}.
\newblock \showarticletitle{{Human3.6M}: Large Scale Datasets and Predictive
  Methods for {3D} Human Sensing in Natural Environments}.
\newblock \bibinfo{journal}{\emph{IEEE Transactions on Pattern Analysis and
  Machine Intelligence, TPAMI}} \bibinfo{volume}{36}, \bibinfo{number}{7}
  (\bibinfo{year}{2014}), \bibinfo{pages}{1325--1339}.
\newblock


\bibitem[\protect\citeauthoryear{J.~Williams and Peng}{J.~Williams and
  Peng}{1991}]%
        {WilliamsPeng1991}
\bibfield{author}{\bibinfo{person}{Ronald J.~Williams} {and}
  \bibinfo{person}{Jing Peng}.} \bibinfo{year}{1991}\natexlab{}.
\newblock \showarticletitle{Function Optimization Using Connectionist
  Reinforcement Learning Algorithms}.
\newblock   \bibinfo{volume}{3} (\bibinfo{date}{09} \bibinfo{year}{1991}),
  \bibinfo{pages}{241--}.
\newblock


\bibitem[\protect\citeauthoryear{Johnson and Everingham}{Johnson and
  Everingham}{2010}]%
        {LSP}
\bibfield{author}{\bibinfo{person}{Sam Johnson} {and} \bibinfo{person}{Mark
  Everingham}.} \bibinfo{year}{2010}\natexlab{}.
\newblock \showarticletitle{Clustered Pose and Nonlinear Appearance Models for
  Human Pose Estimation}. In \bibinfo{booktitle}{\emph{Proceedings of the
  British Machine Vision Conference, BMVC}}. \bibinfo{pages}{12.1--12.11}.
\newblock


\bibitem[\protect\citeauthoryear{Kakade}{Kakade}{2001}]%
        {Kakade2001}
\bibfield{author}{\bibinfo{person}{Sham Kakade}.}
  \bibinfo{year}{2001}\natexlab{}.
\newblock \showarticletitle{A Natural Policy Gradient}. In
  \bibinfo{booktitle}{\emph{Proceedings of the 14th International Conference on
  Neural Information Processing Systems: Natural and Synthetic}}
  \emph{(\bibinfo{series}{NIPS'01})}. \bibinfo{publisher}{MIT Press},
  \bibinfo{address}{Cambridge, MA, USA}, \bibinfo{pages}{1531--1538}.
\newblock
\urldef\tempurl%
\url{http://dl.acm.org/citation.cfm?id=2980539.2980738}
\showURL{%
\tempurl}


\bibitem[\protect\citeauthoryear{Kanazawa, Black, Jacobs, and Malik}{Kanazawa
  et~al\mbox{.}}{2018}]%
        {hmrKanazawa17}
\bibfield{author}{\bibinfo{person}{Angjoo Kanazawa},
  \bibinfo{person}{Michael~J. Black}, \bibinfo{person}{David~W. Jacobs}, {and}
  \bibinfo{person}{Jitendra Malik}.} \bibinfo{year}{2018}\natexlab{}.
\newblock \showarticletitle{End-to-end Recovery of Human Shape and Pose}. In
  \bibinfo{booktitle}{\emph{Computer Vision and Pattern Regognition (CVPR)}}.
\newblock


\bibitem[\protect\citeauthoryear{Lee and Chen}{Lee and Chen}{1985}]%
        {Lee1985}
\bibfield{author}{\bibinfo{person}{H. Lee} {and} \bibinfo{person}{Z. Chen}.}
  \bibinfo{year}{1985}\natexlab{}.
\newblock \showarticletitle{Determination of {3D} human body postures from a
  single view}.
\newblock \bibinfo{journal}{\emph{Computer Vision Graphics and Image
  Processing}} \bibinfo{volume}{30}, \bibinfo{number}{2}
  (\bibinfo{year}{1985}), \bibinfo{pages}{148–--168}.
\newblock


\bibitem[\protect\citeauthoryear{Lee, Kim, and Lee}{Lee et~al\mbox{.}}{2010a}]%
        {Lee2010}
\bibfield{author}{\bibinfo{person}{Yoonsang Lee}, \bibinfo{person}{Sungeun
  Kim}, {and} \bibinfo{person}{Jehee Lee}.} \bibinfo{year}{2010}\natexlab{a}.
\newblock \showarticletitle{Data-driven Biped Control}. In
  \bibinfo{booktitle}{\emph{ACM SIGGRAPH 2010 Papers}}
  \emph{(\bibinfo{series}{SIGGRAPH '10})}. \bibinfo{publisher}{ACM},
  \bibinfo{address}{New York, NY, USA}, Article \bibinfo{articleno}{129},
  \bibinfo{numpages}{8}~pages.
\newblock
\showISBNx{978-1-4503-0210-4}


\bibitem[\protect\citeauthoryear{Lee, Park, Kwon, and Lee}{Lee
  et~al\mbox{.}}{2014}]%
        {Lee2014}
\bibfield{author}{\bibinfo{person}{Yoonsang Lee}, \bibinfo{person}{Moon~Seok
  Park}, \bibinfo{person}{Taesoo Kwon}, {and} \bibinfo{person}{Jehee Lee}.}
  \bibinfo{year}{2014}\natexlab{}.
\newblock \showarticletitle{Locomotion Control for Many-muscle Humanoids}.
\newblock \bibinfo{journal}{\emph{ACM Trans. Graph.}} \bibinfo{volume}{33},
  \bibinfo{number}{6}, Article \bibinfo{articleno}{218} (\bibinfo{date}{Nov.}
  \bibinfo{year}{2014}), \bibinfo{numpages}{11}~pages.
\newblock
\showISSN{0730-0301}


\bibitem[\protect\citeauthoryear{Lee, Wampler, Bernstein, Popovi\'{c}, and
  Popovi\'{c}}{Lee et~al\mbox{.}}{2010b}]%
        {Lee2010MotionFields}
\bibfield{author}{\bibinfo{person}{Yongjoon Lee}, \bibinfo{person}{Kevin
  Wampler}, \bibinfo{person}{Gilbert Bernstein}, \bibinfo{person}{Jovan
  Popovi\'{c}}, {and} \bibinfo{person}{Zoran Popovi\'{c}}.}
  \bibinfo{year}{2010}\natexlab{b}.
\newblock \showarticletitle{Motion Fields for Interactive Character
  Locomotion}. In \bibinfo{booktitle}{\emph{ACM SIGGRAPH Asia 2010 Papers}}
  \emph{(\bibinfo{series}{SIGGRAPH ASIA '10})}. \bibinfo{publisher}{ACM},
  \bibinfo{address}{New York, NY, USA}, Article \bibinfo{articleno}{138},
  \bibinfo{numpages}{8}~pages.
\newblock
\showISBNx{978-1-4503-0439-9}


\bibitem[\protect\citeauthoryear{Levine, Wang, Haraux, Popovi\'{c}, and
  Koltun}{Levine et~al\mbox{.}}{2012}]%
        {2012-ccclde}
\bibfield{author}{\bibinfo{person}{Sergey Levine}, \bibinfo{person}{Jack~M.
  Wang}, \bibinfo{person}{Alexis Haraux}, \bibinfo{person}{Zoran Popovi\'{c}},
  {and} \bibinfo{person}{Vladlen Koltun}.} \bibinfo{year}{2012}\natexlab{}.
\newblock \showarticletitle{Continuous Character Control with Low-Dimensional
  Embeddings}.
\newblock \bibinfo{journal}{\emph{ACM Transactions on Graphics}}
  \bibinfo{volume}{31}, \bibinfo{number}{4} (\bibinfo{year}{2012}),
  \bibinfo{pages}{28}.
\newblock


\bibitem[\protect\citeauthoryear{Lin, Maire, Belongie, Hays, Perona, Ramanan,
  Dollár, and Zitnick}{Lin et~al\mbox{.}}{2014}]%
        {COCO}
\bibfield{author}{\bibinfo{person}{Tsung-Yi Lin}, \bibinfo{person}{Michael
  Maire}, \bibinfo{person}{Serge Belongie}, \bibinfo{person}{James Hays},
  \bibinfo{person}{Pietro Perona}, \bibinfo{person}{Deva Ramanan},
  \bibinfo{person}{Piotr Dollár}, {and} \bibinfo{person}{C.~Lawrence
  Zitnick}.} \bibinfo{year}{2014}\natexlab{}.
\newblock \showarticletitle{Microsoft COCO: Common Objects in Context}. In
  \bibinfo{booktitle}{\emph{European Conference on Computer Vision (ECCV)}}
  (2014-01-01). \bibinfo{address}{Zürich}.
\newblock
\urldef\tempurl%
\url{/se3/wp-content/uploads/2014/09/coco_eccv.pdf, http://mscoco.org}
\showURL{%
\tempurl}


\bibitem[\protect\citeauthoryear{Liu and Hodgins}{Liu and Hodgins}{2017}]%
        {Liu2017}
\bibfield{author}{\bibinfo{person}{Libin Liu} {and} \bibinfo{person}{Jessica
  Hodgins}.} \bibinfo{year}{2017}\natexlab{}.
\newblock \showarticletitle{Learning to Schedule Control Fragments for
  Physics-Based Characters Using Deep Q-Learning}.
\newblock \bibinfo{journal}{\emph{ACM Trans. Graph.}} \bibinfo{volume}{36},
  \bibinfo{number}{3}, Article \bibinfo{articleno}{29} (\bibinfo{date}{June}
  \bibinfo{year}{2017}), \bibinfo{numpages}{14}~pages.
\newblock
\showISSN{0730-0301}


\bibitem[\protect\citeauthoryear{Liu, van~de Panne, and Yin}{Liu
  et~al\mbox{.}}{2016}]%
        {2016-TOG-controlGraphs}
\bibfield{author}{\bibinfo{person}{Libin Liu}, \bibinfo{person}{Michiel van~de
  Panne}, {and} \bibinfo{person}{KangKang Yin}.}
  \bibinfo{year}{2016}\natexlab{}.
\newblock \showarticletitle{Guided Learning of Control Graphs for Physics-Based
  Characters}.
\newblock \bibinfo{journal}{\emph{ACM Transactions on Graphics}}
  \bibinfo{volume}{35}, \bibinfo{number}{3} (\bibinfo{year}{2016}).
\newblock


\bibitem[\protect\citeauthoryear{Liu, Yin, van~de Panne, Shao, and Xu}{Liu
  et~al\mbox{.}}{2010}]%
        {2010-TOG-sampControl}
\bibfield{author}{\bibinfo{person}{Libin Liu}, \bibinfo{person}{KangKang Yin},
  \bibinfo{person}{Michiel van~de Panne}, \bibinfo{person}{Tianjia Shao}, {and}
  \bibinfo{person}{Weiwei Xu}.} \bibinfo{year}{2010}\natexlab{}.
\newblock \showarticletitle{Sampling-based Contact-rich Motion Control}.
\newblock \bibinfo{journal}{\emph{ACM Transctions on Graphics}}
  \bibinfo{volume}{29}, \bibinfo{number}{4} (\bibinfo{year}{2010}),
  \bibinfo{pages}{Article 128}.
\newblock


\bibitem[\protect\citeauthoryear{Loper, Mahmood, Romero, Pons-Moll, and
  Black}{Loper et~al\mbox{.}}{2015}]%
        {SMPL}
\bibfield{author}{\bibinfo{person}{Matthew Loper}, \bibinfo{person}{Naureen
  Mahmood}, \bibinfo{person}{Javier Romero}, \bibinfo{person}{Gerard
  Pons-Moll}, {and} \bibinfo{person}{Michael~J. Black}.}
  \bibinfo{year}{2015}\natexlab{}.
\newblock \showarticletitle{{SMPL}: A Skinned Multi-Person Linear Model}.
\newblock \bibinfo{journal}{\emph{ACM Trans. Graphics (Proc. SIGGRAPH Asia)}}
  \bibinfo{volume}{34}, \bibinfo{number}{6} (\bibinfo{date}{Oct.}
  \bibinfo{year}{2015}), \bibinfo{pages}{248:1--248:16}.
\newblock


\bibitem[\protect\citeauthoryear{Mehta, Sridhar, Sotnychenko, Rhodin, Shafiei,
  Seidel, Xu, Casas, and Theobalt}{Mehta et~al\mbox{.}}{2017}]%
        {VNect}
\bibfield{author}{\bibinfo{person}{Dushyant Mehta}, \bibinfo{person}{Srinath
  Sridhar}, \bibinfo{person}{Oleksandr Sotnychenko}, \bibinfo{person}{Helge
  Rhodin}, \bibinfo{person}{Mohammad Shafiei}, \bibinfo{person}{Hans-Peter
  Seidel}, \bibinfo{person}{Weipeng Xu}, \bibinfo{person}{Dan Casas}, {and}
  \bibinfo{person}{Christian Theobalt}.} \bibinfo{year}{2017}\natexlab{}.
\newblock \showarticletitle{VNect: Real-time 3D Human Pose Estimation with a
  Single RGB Camera}.
\newblock \bibinfo{journal}{\emph{ACM Transactions on Graphics (TOG) -
  Proceedings of ACM SIGGRAPH}}  \bibinfo{volume}{36} (\bibinfo{date}{July}
  \bibinfo{year}{2017}), 14.
\newblock
\urldef\tempurl%
\url{http://gvv.mpi-inf.mpg.de/projects/VNect/}
\showURL{%
\tempurl}


\bibitem[\protect\citeauthoryear{Merel, Tassa, TB, Srinivasan, Lemmon, Wang,
  Wayne, and Heess}{Merel et~al\mbox{.}}{2017}]%
        {MerelTTSLWWH17}
\bibfield{author}{\bibinfo{person}{Josh Merel}, \bibinfo{person}{Yuval Tassa},
  \bibinfo{person}{Dhruva TB}, \bibinfo{person}{Sriram Srinivasan},
  \bibinfo{person}{Jay Lemmon}, \bibinfo{person}{Ziyu Wang},
  \bibinfo{person}{Greg Wayne}, {and} \bibinfo{person}{Nicolas Heess}.}
  \bibinfo{year}{2017}\natexlab{}.
\newblock \showarticletitle{Learning human behaviors from motion capture by
  adversarial imitation}.
\newblock \bibinfo{journal}{\emph{CoRR}}  \bibinfo{volume}{abs/1707.02201}
  (\bibinfo{year}{2017}).
\newblock
\showeprint[arxiv]{1707.02201}


\bibitem[\protect\citeauthoryear{Newell, Yang, and Deng}{Newell
  et~al\mbox{.}}{2016}]%
        {hourglass}
\bibfield{author}{\bibinfo{person}{Alejandro Newell}, \bibinfo{person}{Kaiyu
  Yang}, {and} \bibinfo{person}{Jia Deng}.} \bibinfo{year}{2016}\natexlab{}.
\newblock \showarticletitle{Stacked hourglass networks for human pose
  estimation}. In \bibinfo{booktitle}{\emph{European Conference on Computer
  Vision, ECCV}}. \bibinfo{pages}{483--499}.
\newblock


\bibitem[\protect\citeauthoryear{Pavlakos, Zhou, Derpanis, and
  Daniilidis}{Pavlakos et~al\mbox{.}}{2017}]%
        {Pavlakos}
\bibfield{author}{\bibinfo{person}{Georgios Pavlakos}, \bibinfo{person}{Xiaowei
  Zhou}, \bibinfo{person}{Konstantinos~G Derpanis}, {and}
  \bibinfo{person}{Kostas Daniilidis}.} \bibinfo{year}{2017}\natexlab{}.
\newblock \showarticletitle{Coarse-to-Fine Volumetric Prediction for
  Single-Image 3{D} Human Pose}. In \bibinfo{booktitle}{\emph{IEEE Conference
  on Computer Vision and Pattern Recognition, CVPR}}.
\newblock


\bibitem[\protect\citeauthoryear{Peng, Abbeel, Levine, and van~de Panne}{Peng
  et~al\mbox{.}}{2018}]%
        {2018-TOG-DeepMimic}
\bibfield{author}{\bibinfo{person}{Xue~Bin Peng}, \bibinfo{person}{Pieter
  Abbeel}, \bibinfo{person}{Sergey Levine}, {and} \bibinfo{person}{Michiel
  van~de Panne}.} \bibinfo{year}{2018}\natexlab{}.
\newblock \showarticletitle{DeepMimic: Example-Guided Deep Reinforcement
  Learning of Physics-Based Character Skills}.
\newblock \bibinfo{journal}{\emph{ACM Transactions on Graphics (Proc. SIGGRAPH
  2018 - to appear)}} \bibinfo{volume}{37}, \bibinfo{number}{4}
  (\bibinfo{year}{2018}).
\newblock


\bibitem[\protect\citeauthoryear{Peng, Berseth, and van~de Panne}{Peng
  et~al\mbox{.}}{2015}]%
        {2015-TOG-terrainRL}
\bibfield{author}{\bibinfo{person}{Xue~Bin Peng}, \bibinfo{person}{Glen
  Berseth}, {and} \bibinfo{person}{Michiel van~de Panne}.}
  \bibinfo{year}{2015}\natexlab{}.
\newblock \showarticletitle{Dynamic Terrain Traversal Skills Using
  Reinforcement Learning}.
\newblock \bibinfo{journal}{\emph{ACM Trans. Graph.}} \bibinfo{volume}{34},
  \bibinfo{number}{4}, Article \bibinfo{articleno}{80} (\bibinfo{date}{July}
  \bibinfo{year}{2015}), \bibinfo{numpages}{11}~pages.
\newblock
\showISSN{0730-0301}


\bibitem[\protect\citeauthoryear{Peng, Berseth, and van~de Panne}{Peng
  et~al\mbox{.}}{2016}]%
        {2016-TOG-deepRL}
\bibfield{author}{\bibinfo{person}{Xue~Bin Peng}, \bibinfo{person}{Glen
  Berseth}, {and} \bibinfo{person}{Michiel van~de Panne}.}
  \bibinfo{year}{2016}\natexlab{}.
\newblock \showarticletitle{Terrain-Adaptive Locomotion Skills Using Deep
  Reinforcement Learning}.
\newblock \bibinfo{journal}{\emph{ACM Transactions on Graphics (Proc. SIGGRAPH
  2016)}} \bibinfo{volume}{35}, \bibinfo{number}{4} (\bibinfo{year}{2016}).
\newblock


\bibitem[\protect\citeauthoryear{Rajeswaran, Kumar, Gupta, Schulman, Todorov,
  and Levine}{Rajeswaran et~al\mbox{.}}{2017}]%
        {Rajeswaran2017}
\bibfield{author}{\bibinfo{person}{Aravind Rajeswaran}, \bibinfo{person}{Vikash
  Kumar}, \bibinfo{person}{Abhishek Gupta}, \bibinfo{person}{John Schulman},
  \bibinfo{person}{Emanuel Todorov}, {and} \bibinfo{person}{Sergey Levine}.}
  \bibinfo{year}{2017}\natexlab{}.
\newblock \showarticletitle{Learning Complex Dexterous Manipulation with Deep
  Reinforcement Learning and Demonstrations}.
\newblock \bibinfo{journal}{\emph{CoRR}}  \bibinfo{volume}{abs/1709.10087}
  (\bibinfo{year}{2017}).
\newblock
\showeprint[arxiv]{1709.10087}


\bibitem[\protect\citeauthoryear{Rogez and Schmid}{Rogez and Schmid}{2016}]%
        {RogezMocap}
\bibfield{author}{\bibinfo{person}{Gregory Rogez} {and}
  \bibinfo{person}{Cordelia Schmid}.} \bibinfo{year}{2016}\natexlab{}.
\newblock \showarticletitle{MoCap-guided Data Augmentation for 3D Pose
  Estimation in the Wild}. In \bibinfo{booktitle}{\emph{Advances in Neural
  Information Processing Systems, (NIPS)}}.
\newblock


\bibitem[\protect\citeauthoryear{Schulman, Levine, Moritz, Jordan, and
  Abbeel}{Schulman et~al\mbox{.}}{2015a}]%
        {SchulmanLMJA15}
\bibfield{author}{\bibinfo{person}{John Schulman}, \bibinfo{person}{Sergey
  Levine}, \bibinfo{person}{Philipp Moritz}, \bibinfo{person}{Michael~I.
  Jordan}, {and} \bibinfo{person}{Pieter Abbeel}.}
  \bibinfo{year}{2015}\natexlab{a}.
\newblock \showarticletitle{Trust Region Policy Optimization}.
\newblock \bibinfo{journal}{\emph{CoRR}}  \bibinfo{volume}{abs/1502.05477}
  (\bibinfo{year}{2015}).
\newblock
\showeprint[arxiv]{1502.05477}


\bibitem[\protect\citeauthoryear{Schulman, Moritz, Levine, Jordan, and
  Abbeel}{Schulman et~al\mbox{.}}{2015b}]%
        {SchulmanMLJA15}
\bibfield{author}{\bibinfo{person}{John Schulman}, \bibinfo{person}{Philipp
  Moritz}, \bibinfo{person}{Sergey Levine}, \bibinfo{person}{Michael~I.
  Jordan}, {and} \bibinfo{person}{Pieter Abbeel}.}
  \bibinfo{year}{2015}\natexlab{b}.
\newblock \showarticletitle{High-Dimensional Continuous Control Using
  Generalized Advantage Estimation}.
\newblock \bibinfo{journal}{\emph{CoRR}}  \bibinfo{volume}{abs/1506.02438}
  (\bibinfo{year}{2015}).
\newblock
\showeprint[arxiv]{1506.02438}


\bibitem[\protect\citeauthoryear{Schulman, Wolski, Dhariwal, Radford, and
  Klimov}{Schulman et~al\mbox{.}}{2017}]%
        {PPO17}
\bibfield{author}{\bibinfo{person}{John Schulman}, \bibinfo{person}{Filip
  Wolski}, \bibinfo{person}{Prafulla Dhariwal}, \bibinfo{person}{Alec Radford},
  {and} \bibinfo{person}{Oleg Klimov}.} \bibinfo{year}{2017}\natexlab{}.
\newblock \showarticletitle{Proximal Policy Optimization Algorithms}.
\newblock \bibinfo{journal}{\emph{CoRR}}  \bibinfo{volume}{abs/1707.06347}
  (\bibinfo{year}{2017}).
\newblock
\showeprint[arxiv]{1707.06347}


\bibitem[\protect\citeauthoryear{Sermanet, Lynch, Hsu, and Levine}{Sermanet
  et~al\mbox{.}}{2017}]%
        {SermanetLHL17}
\bibfield{author}{\bibinfo{person}{Pierre Sermanet}, \bibinfo{person}{Corey
  Lynch}, \bibinfo{person}{Jasmine Hsu}, {and} \bibinfo{person}{Sergey
  Levine}.} \bibinfo{year}{2017}\natexlab{}.
\newblock \showarticletitle{Time-Contrastive Networks: Self-Supervised Learning
  from Multi-View Observation}.
\newblock \bibinfo{journal}{\emph{CoRR}}  \bibinfo{volume}{abs/1704.06888}
  (\bibinfo{year}{2017}).
\newblock
\showeprint[arxiv]{1704.06888}
\urldef\tempurl%
\url{http://arxiv.org/abs/1704.06888}
\showURL{%
\tempurl}


\bibitem[\protect\citeauthoryear{SFU}{SFU}{2018}]%
        {SFUMocap}
\bibfield{author}{\bibinfo{person}{SFU}.} \bibinfo{year}{2018}\natexlab{}.
\newblock \bibinfo{title}{SFU Motion Capture Database}.
\newblock   (\bibinfo{year}{2018}).
\newblock
\newblock
\shownote{http://mocap.cs.sfu.ca.}


\bibitem[\protect\citeauthoryear{Sutton, Mcallester, Singh, and Mansour}{Sutton
  et~al\mbox{.}}{2001}]%
        {sutton2001policy}
\bibfield{author}{\bibinfo{person}{R. Sutton}, \bibinfo{person}{D. Mcallester},
  \bibinfo{person}{S. Singh}, {and} \bibinfo{person}{Y. Mansour}.}
  \bibinfo{year}{2001}\natexlab{}.
\newblock \bibinfo{title}{Policy Gradient Methods for Reinforcement Learning
  with Function Approximation}.
\newblock   (\bibinfo{year}{2001}), \bibinfo{numpages}{1057--1063}~pages.
\newblock


\bibitem[\protect\citeauthoryear{Sutton and Barto}{Sutton and Barto}{1998}]%
        {Sutton1998}
\bibfield{author}{\bibinfo{person}{Richard~S. Sutton} {and}
  \bibinfo{person}{Andrew~G. Barto}.} \bibinfo{year}{1998}\natexlab{}.
\newblock \bibinfo{booktitle}{\emph{Introduction to Reinforcement Learning}
  (\bibinfo{edition}{1st} ed.)}.
\newblock \bibinfo{publisher}{MIT Press}, \bibinfo{address}{Cambridge, MA,
  USA}.
\newblock
\showISBNx{0262193981}


\bibitem[\protect\citeauthoryear{Taylor}{Taylor}{2000}]%
        {Taylor:2000}
\bibfield{author}{\bibinfo{person}{C. Taylor}.}
  \bibinfo{year}{2000}\natexlab{}.
\newblock \showarticletitle{Reconstruction of articulated objects from point
  correspondences in single uncalibrated image}.
\newblock \bibinfo{journal}{\emph{Computer Vision and Image Understanding,
  CVIU}} \bibinfo{volume}{80}, \bibinfo{number}{10} (\bibinfo{year}{2000}),
  \bibinfo{pages}{349--363}.
\newblock


\bibitem[\protect\citeauthoryear{Teh, Bapst, Czarnecki, Quan, Kirkpatrick,
  Hadsell, Heess, and Pascanu}{Teh et~al\mbox{.}}{2017}]%
        {TehBCQKHHP17}
\bibfield{author}{\bibinfo{person}{Yee~Whye Teh}, \bibinfo{person}{Victor
  Bapst}, \bibinfo{person}{Wojciech~Marian Czarnecki}, \bibinfo{person}{John
  Quan}, \bibinfo{person}{James Kirkpatrick}, \bibinfo{person}{Raia Hadsell},
  \bibinfo{person}{Nicolas Heess}, {and} \bibinfo{person}{Razvan Pascanu}.}
  \bibinfo{year}{2017}\natexlab{}.
\newblock \showarticletitle{Distral: Robust Multitask Reinforcement Learning}.
\newblock \bibinfo{journal}{\emph{CoRR}}  \bibinfo{volume}{abs/1707.04175}
  (\bibinfo{year}{2017}).
\newblock
\showeprint[arxiv]{1707.04175}


\bibitem[\protect\citeauthoryear{Tekin, Rozantsev, Lepetit, and Fua}{Tekin
  et~al\mbox{.}}{2016}]%
        {Tekin:2015}
\bibfield{author}{\bibinfo{person}{Bugra Tekin}, \bibinfo{person}{Artem
  Rozantsev}, \bibinfo{person}{Vincent Lepetit}, {and} \bibinfo{person}{Pascal
  Fua}.} \bibinfo{year}{2016}\natexlab{}.
\newblock \showarticletitle{Direct Prediction of {3D} Body Poses from Motion
  Compensated Sequences}. In \bibinfo{booktitle}{\emph{IEEE Conference on
  Computer Vision and Pattern Recognition, CVPR}}. \bibinfo{pages}{991--1000}.
\newblock


\bibitem[\protect\citeauthoryear{Tompson, Jain, LeCun, and Bregler}{Tompson
  et~al\mbox{.}}{2014}]%
        {tompson2014joint}
\bibfield{author}{\bibinfo{person}{Jonathan~J Tompson}, \bibinfo{person}{Arjun
  Jain}, \bibinfo{person}{Yann LeCun}, {and} \bibinfo{person}{Christoph
  Bregler}.} \bibinfo{year}{2014}\natexlab{}.
\newblock \showarticletitle{Joint training of a convolutional network and a
  graphical model for human pose estimation}. In
  \bibinfo{booktitle}{\emph{Advances in neural information processing
  systems}}. \bibinfo{pages}{1799--1807}.
\newblock


\bibitem[\protect\citeauthoryear{Toshev and Szegedy}{Toshev and
  Szegedy}{2014}]%
        {Toshev:2014}
\bibfield{author}{\bibinfo{person}{Alexander Toshev} {and}
  \bibinfo{person}{Christian Szegedy}.} \bibinfo{year}{2014}\natexlab{}.
\newblock \showarticletitle{Deep{P}ose: Human Pose Estimation via Deep Neural
  Networks}. In \bibinfo{booktitle}{\emph{IEEE Conference on Computer Vision
  and Pattern Recognition, CVPR}}. \bibinfo{pages}{1653--1660}.
\newblock


\bibitem[\protect\citeauthoryear{Tung, Tung, Yumer, and Fragkiadaki}{Tung
  et~al\mbox{.}}{2017}]%
        {tung2017self}
\bibfield{author}{\bibinfo{person}{Hsiao-Yu Tung}, \bibinfo{person}{Hsiao-Wei
  Tung}, \bibinfo{person}{Ersin Yumer}, {and} \bibinfo{person}{Katerina
  Fragkiadaki}.} \bibinfo{year}{2017}\natexlab{}.
\newblock \showarticletitle{Self-supervised Learning of Motion Capture}. In
  \bibinfo{booktitle}{\emph{Advances in Neural Information Processing
  Systems}}. \bibinfo{pages}{5242--5252}.
\newblock


\bibitem[\protect\citeauthoryear{Vondrak, Sigal, Hodgins, and Jenkins}{Vondrak
  et~al\mbox{.}}{2012}]%
        {Vondrak2012}
\bibfield{author}{\bibinfo{person}{Marek Vondrak}, \bibinfo{person}{Leonid
  Sigal}, \bibinfo{person}{Jessica Hodgins}, {and} \bibinfo{person}{Odest
  Jenkins}.} \bibinfo{year}{2012}\natexlab{}.
\newblock \showarticletitle{Video-based 3D Motion Capture Through Biped
  Control}.
\newblock \bibinfo{journal}{\emph{ACM Trans. Graph.}} \bibinfo{volume}{31},
  \bibinfo{number}{4}, Article \bibinfo{articleno}{27} (\bibinfo{date}{July}
  \bibinfo{year}{2012}), \bibinfo{numpages}{12}~pages.
\newblock
\showISSN{0730-0301}
\urldef\tempurl%
\url{https://doi.org/10.1145/2185520.2185523}
\showDOI{\tempurl}


\bibitem[\protect\citeauthoryear{Wampler, Popovi\'{c}, and Popovi\'{c}}{Wampler
  et~al\mbox{.}}{2014}]%
        {Wampler2014}
\bibfield{author}{\bibinfo{person}{Kevin Wampler}, \bibinfo{person}{Zoran
  Popovi\'{c}}, {and} \bibinfo{person}{Jovan Popovi\'{c}}.}
  \bibinfo{year}{2014}\natexlab{}.
\newblock \showarticletitle{Generalizing Locomotion Style to New Animals with
  Inverse Optimal Regression}.
\newblock \bibinfo{journal}{\emph{ACM Trans. Graph.}} \bibinfo{volume}{33},
  \bibinfo{number}{4}, Article \bibinfo{articleno}{49} (\bibinfo{date}{July}
  \bibinfo{year}{2014}), \bibinfo{numpages}{11}~pages.
\newblock
\showISSN{0730-0301}


\bibitem[\protect\citeauthoryear{Wang, Fleet, and Hertzmann}{Wang
  et~al\mbox{.}}{2010}]%
        {Wang2010}
\bibfield{author}{\bibinfo{person}{Jack~M. Wang}, \bibinfo{person}{David~J.
  Fleet}, {and} \bibinfo{person}{Aaron Hertzmann}.}
  \bibinfo{year}{2010}\natexlab{}.
\newblock \showarticletitle{Optimizing Walking Controllers for Uncertain Inputs
  and Environments}. In \bibinfo{booktitle}{\emph{ACM SIGGRAPH 2010 Papers}}
  \emph{(\bibinfo{series}{SIGGRAPH '10})}. \bibinfo{publisher}{ACM},
  \bibinfo{address}{New York, NY, USA}, Article \bibinfo{articleno}{73},
  \bibinfo{numpages}{8}~pages.
\newblock
\showISBNx{978-1-4503-0210-4}
\urldef\tempurl%
\url{https://doi.org/10.1145/1833349.1778810}
\showDOI{\tempurl}


\bibitem[\protect\citeauthoryear{Wang, Hamner, Delp, and Koltun}{Wang
  et~al\mbox{.}}{2012}]%
        {Wang2012}
\bibfield{author}{\bibinfo{person}{Jack~M. Wang}, \bibinfo{person}{Samuel~R.
  Hamner}, \bibinfo{person}{Scott~L. Delp}, {and} \bibinfo{person}{Vladlen
  Koltun}.} \bibinfo{year}{2012}\natexlab{}.
\newblock \showarticletitle{Optimizing Locomotion Controllers Using
  Biologically-based Actuators and Objectives}.
\newblock \bibinfo{journal}{\emph{ACM Trans. Graph.}} \bibinfo{volume}{31},
  \bibinfo{number}{4}, Article \bibinfo{articleno}{25} (\bibinfo{date}{July}
  \bibinfo{year}{2012}), \bibinfo{numpages}{11}~pages.
\newblock
\showISSN{0730-0301}
\urldef\tempurl%
\url{https://doi.org/10.1145/2185520.2185521}
\showDOI{\tempurl}


\bibitem[\protect\citeauthoryear{Wei, Ramakrishna, Kanade, and Sheikh}{Wei
  et~al\mbox{.}}{2016}]%
        {Wei:CVPR:2016}
\bibfield{author}{\bibinfo{person}{Shih-En Wei}, \bibinfo{person}{Varun
  Ramakrishna}, \bibinfo{person}{Takeo Kanade}, {and} \bibinfo{person}{Yaser
  Sheikh}.} \bibinfo{year}{2016}\natexlab{}.
\newblock \showarticletitle{Convolutional Pose Machines}. In
  \bibinfo{booktitle}{\emph{IEEE Conference on Computer Vision and Pattern
  Recognition, CVPR}}. \bibinfo{pages}{4724--4732}.
\newblock


\bibitem[\protect\citeauthoryear{Won, Park, Kim, and Lee}{Won
  et~al\mbox{.}}{2017}]%
        {Won2017}
\bibfield{author}{\bibinfo{person}{Jungdam Won}, \bibinfo{person}{Jongho Park},
  \bibinfo{person}{Kwanyu Kim}, {and} \bibinfo{person}{Jehee Lee}.}
  \bibinfo{year}{2017}\natexlab{}.
\newblock \showarticletitle{How to Train Your Dragon: Example-guided Control of
  Flapping Flight}.
\newblock \bibinfo{journal}{\emph{ACM Trans. Graph.}} \bibinfo{volume}{36},
  \bibinfo{number}{6}, Article \bibinfo{articleno}{198} (\bibinfo{date}{Nov.}
  \bibinfo{year}{2017}), \bibinfo{numpages}{13}~pages.
\newblock
\showISSN{0730-0301}


\bibitem[\protect\citeauthoryear{Xu, Chatterjee, Zollhoefer, Rhodin, Mehta,
  Seidel, and Theobalt}{Xu et~al\mbox{.}}{2018}]%
        {MonoPerfCap_SIGGRAPH2018}
\bibfield{author}{\bibinfo{person}{Weipeng Xu}, \bibinfo{person}{Avishek
  Chatterjee}, \bibinfo{person}{Michael Zollhoefer}, \bibinfo{person}{Helge
  Rhodin}, \bibinfo{person}{Dushyant Mehta}, \bibinfo{person}{Hans-Peter
  Seidel}, {and} \bibinfo{person}{Christian Theobalt}.}
  \bibinfo{year}{2018}\natexlab{}.
\newblock \showarticletitle{MonoPerfCap: Human Performance Capture from
  Monocular Video}.
\newblock \bibinfo{journal}{\emph{ACM Transactions on Graphics}}
  (\bibinfo{year}{2018}).
\newblock
\urldef\tempurl%
\url{http://gvv.mpi-inf.mpg.de/projects/wxu/MonoPerfCap}
\showURL{%
\tempurl}


\bibitem[\protect\citeauthoryear{Yu, Finn, Xie, Dasari, Zhang, Abbeel, and
  Levine}{Yu et~al\mbox{.}}{2018a}]%
        {Yu2018}
\bibfield{author}{\bibinfo{person}{Tianhe Yu}, \bibinfo{person}{Chelsea Finn},
  \bibinfo{person}{Annie Xie}, \bibinfo{person}{Sudeep Dasari},
  \bibinfo{person}{Tianhao Zhang}, \bibinfo{person}{Pieter Abbeel}, {and}
  \bibinfo{person}{Sergey Levine}.} \bibinfo{year}{2018}\natexlab{a}.
\newblock \showarticletitle{One-Shot Imitation from Observing Humans via
  Domain-Adaptive Meta-Learning}.
\newblock \bibinfo{journal}{\emph{CoRR}}  \bibinfo{volume}{abs/1802.01557}
  (\bibinfo{year}{2018}).
\newblock
\showeprint[arxiv]{1802.01557}
\urldef\tempurl%
\url{http://arxiv.org/abs/1802.01557}
\showURL{%
\tempurl}


\bibitem[\protect\citeauthoryear{Yu, Turk, and Liu}{Yu et~al\mbox{.}}{2018b}]%
        {YuMSL2018}
\bibfield{author}{\bibinfo{person}{Wenhao Yu}, \bibinfo{person}{Greg Turk},
  {and} \bibinfo{person}{C.~Karen Liu}.} \bibinfo{year}{2018}\natexlab{b}.
\newblock \showarticletitle{Learning Symmetry and Low-energy Locomotion}.
\newblock \bibinfo{journal}{\emph{CoRR}}  \bibinfo{volume}{abs/1801.08093}
  (\bibinfo{year}{2018}).
\newblock
\showeprint[arxiv]{1801.08093}
\urldef\tempurl%
\url{http://arxiv.org/abs/1801.08093}
\showURL{%
\tempurl}


\bibitem[\protect\citeauthoryear{Zhou, Huang, Sun, Xue, and Wei}{Zhou
  et~al\mbox{.}}{2017}]%
        {Xingyi2017}
\bibfield{author}{\bibinfo{person}{Xingyi Zhou}, \bibinfo{person}{Qixing
  Huang}, \bibinfo{person}{Xiao Sun}, \bibinfo{person}{Xiangyang Xue}, {and}
  \bibinfo{person}{Yichen Wei}.} \bibinfo{year}{2017}\natexlab{}.
\newblock \showarticletitle{Weakly-supervised Transfer for 3D Human Pose
  Estimation in the Wild}. In \bibinfo{booktitle}{\emph{IEEE International
  Conference on Computer Vision, ICCV}}.
\newblock


\bibitem[\protect\citeauthoryear{Zhou, Sun, Zhang, Liang, and Wei}{Zhou
  et~al\mbox{.}}{2016}]%
        {Xingyi2016}
\bibfield{author}{\bibinfo{person}{Xingyi Zhou}, \bibinfo{person}{Xiao Sun},
  \bibinfo{person}{Wei Zhang}, \bibinfo{person}{Shuang Liang}, {and}
  \bibinfo{person}{Yichen Wei}.} \bibinfo{year}{2016}\natexlab{}.
\newblock \showarticletitle{Deep Kinematic Pose Regression}. In
  \bibinfo{booktitle}{\emph{ECCV Workshop on Geometry Meets Deep Learning}}.
  \bibinfo{pages}{186--201}.
\newblock


\bibitem[\protect\citeauthoryear{{Zhou}, {Zhu}, {Leonardos}, {Derpanis}, and
  {Daniilidis}}{{Zhou} et~al\mbox{.}}{2015}]%
        {Zhou:2015b}
\bibfield{author}{\bibinfo{person}{X. {Zhou}}, \bibinfo{person}{M. {Zhu}},
  \bibinfo{person}{S. {Leonardos}}, \bibinfo{person}{K. {Derpanis}}, {and}
  \bibinfo{person}{K. {Daniilidis}}.} \bibinfo{year}{2015}\natexlab{}.
\newblock \showarticletitle{Sparse Representation for {3D} Shape Estimation:
  {A} Convex Relaxation Approach}. In \bibinfo{booktitle}{\emph{IEEE Conference
  on Computer Vision and Pattern Recognition, CVPR}}.
  \bibinfo{pages}{4447--4455}.
\newblock


\bibitem[\protect\citeauthoryear{{Zhou}, {Zhu}, {Leonardos}, {Derpanis}, and
  {Daniilidis}}{{Zhou} et~al\mbox{.}}{2016}]%
        {Zhou:2016a}
\bibfield{author}{\bibinfo{person}{X. {Zhou}}, \bibinfo{person}{M. {Zhu}},
  \bibinfo{person}{S. {Leonardos}}, \bibinfo{person}{K. {Derpanis}}, {and}
  \bibinfo{person}{K. {Daniilidis}}.} \bibinfo{year}{2016}\natexlab{}.
\newblock \showarticletitle{Sparseness Meets Deepness: {3D} Human Pose
  Estimation from Monocular Video}. In \bibinfo{booktitle}{\emph{IEEE
  Conference on Computer Vision and Pattern Recognition, CVPR}}.
  \bibinfo{pages}{4966--4975}.
\newblock


\end{thebibliography}

\newpage
\clearpage

\section*{Supplementary Material}

\appendix

\begin{table}[h]
\centering
\caption{Summary of Notations}
\label{tab:notation}
\begin{tabular}{ll}
\hline
Notation       & Definition                                                                          \\ \hline
$\hat{x}_t$    & 2D joint prediction at frame $t$                                                    \\
$\hat{q}_t$    & 3D pose prediction at frame $t$                                                     \\
$z_t$          & HMR embedding of the image frame at $t$                                             \\
$q(z_t)$       & 3D pose as a function of the embedding                                                \\
$F_{j}(\cdot)$ & Forward kinematics function that computes
\\&the 3D position of joint $j$ \\
$q^*_t$        & Final 3D reference pose after motion reconstruction 
\\
$s_t$           & state of the simulated character at timestep $t$
\\
$a_t$           & action
\\
$r_t$           & reward
\\
$R_t$           & return starting at timestep $t$, $\sum_{l = 0}^{T - t} \gamma^l r_{t + l}$
\\
$\mathcal{A}_t$           & advantage at timestep $t$, $R_t - V(s_t)$
\\
$\tau$          & a simulated trajectory $(s_0, a_0, s_1, ..., a_{T - 1}, s_T),$
\\
$\pi_\theta(a|s)$      & policy with parameters $\theta$
\\
$\rho_\omega(s_0)$      & initial state distribution with parameters $\omega$ \\
$\alpha_\pi$        & policy stepsize \\
$\alpha_V$          & value function stepsize \\
$\alpha_\rho$       & initial state distribution stepsize \\
 \hline
\end{tabular}
\end{table}

\section{Learning Algorithm}
When training with adaptive state initialization, the multi-agent objective is given by:
\[J(\theta, \omega) = \mathbb{E}_{\tau \sim p_{\theta, \omega}(\tau)} \left[\sum_{t = 0}^T \gamma^t r_t \right]\]
\[= \int_\tau \left( \rho_\omega(s_0) \prod_{t = 0}^{T - 1} p(s_{t + 1} | s_t, a_t) \pi_\theta(a_t | s_t) \right) \left(\sum_{t = 0}^T \gamma^t r_t \right) d\tau .\]
The policy gradient of the initial state distribution $\rho_\omega(s_0)$ can then be as follows:
\[\triangledown_\omega J(\theta, \omega) = \int_\tau \left( \triangledown_\omega \rho_\omega(s_0) \prod_{t = 0}^{T - 1} p(s_{t + 1} | s_t, a_t) \pi_\theta(a_t | s_t) \right) \left(\sum_{t = 0}^T \gamma^t r_t \right) d\tau \]
\[ = \int_\tau \left( \rho_\omega(s_0) \frac{\triangledown_\omega \rho_\omega(s_0)}{\rho_\omega(s_0)} \prod_{t = 0}^{T - 1} p(s_{t + 1} | s_t, a_t) \pi_\theta(a_t | s_t) \right) \left(\sum_{t = 0}^T \gamma^t r_t \right) d\tau \]
\[ = \int_\tau \left(\triangledown_\omega \mathrm{log} \left(\rho_\omega(s_0) \right) \rho_\omega(s_0) \prod_{t = 0}^{T - 1} p(s_{t + 1} | s_t, a_t) \pi_\theta(a_t | s_t) \right) \left(\sum_{t = 0}^T \gamma^t r_t \right) d\tau \]
\[ = \mathbb{E}_{\tau \sim p_{\theta, \omega}(\tau)} \left[\triangledown_\omega \mathrm{log} \left(\rho_\omega(s_0) \right) \sum_{t = 0}^T \gamma^t r_t \right]. \]

Algorithm~\ref{alg:ASI} provides an overview of the training process. Policy updates are performed after a batch of $4096$ samples has been collected, and minibatches of size $256$ are then sampled from the data for each gradient step. The initial state distribution is updated using batches of $2000$ episodes. Only one gradient step is calculated per batch when updating the initial state distribution. SGD with a stepsize of $\alpha_\pi = 2.5 \times 10^{-6}$ and momentum of 0.9 is used for the policy network, a stepsize of $\alpha_V = 0.01$ is used for the value network, and a stepsize of $\alpha_\rho = 0.001$ is used for the initial state distribution. The discount factor $\gamma$ is set to $0.95$, and $\lambda$ is set to $0.95$ for TD($\lambda$) and GAE($\lambda$). PPO with a clipping threshold of $0.2$ is used for policy updates.

\begin{algorithm}[t!]
\caption{Policy Gradient with Adaptive State Initialization}
\label{alg:ASI}
\begin{algorithmic}[1]
\STATE{$\theta \leftarrow$ initialize policy parameters}
\STATE{$\psi \leftarrow$ initialize value function parameters}
\STATE{$\omega \leftarrow$ initialize initial state distribution parameters}
\item[]
\WHILE{not done}
	\STATE{$s_0 \leftarrow$ sample initial state from $\rho_\omega(s_0)$}
    \STATE{Simulate rollout $(s_0, a_0, r_0, s_1,...,s_{T + 1})$ starting in state $s_0$}
    
    \item[]
	\FOR{each timestep $t$}
		\STATE{$R_t \leftarrow $ compute return using $V_\psi$ and TD($\lambda$)}
		\STATE{$\mathcal{A}_t \leftarrow $ compute advantage using $V_\psi$ and GAE($\lambda$)}
	\ENDFOR
	
    \item[]
	\STATE{Update value function:}
    \STATE{$\psi \leftarrow \psi - \alpha_V \left( \frac{1}{T} \sum_t \triangledown_{\psi} V_\psi(s_t) (R_t - V(s_t))     \right)$}
    
    \item[]
	\STATE{Update policy:}
    \STATE{$\theta \leftarrow \theta + \alpha_\pi \left( \frac{1}{T} \sum_t \triangledown_{\theta} \mathrm{log}\left(\pi_\theta(a_t | s_t)\right) \mathcal{A}_t     \right)$}
		
    \item[]
	\STATE{Update initial state distribution:}
    \STATE{$\omega \leftarrow \omega + \alpha_\rho \triangledown_{\omega} \mathrm{log}\left(\rho_\omega(s_0)\right) R_0$}
\ENDWHILE
\end{algorithmic}
\end{algorithm}

\begin{figure*}[t]
	\centering
     \subfigure{   \includegraphics[width=0.5\columnwidth]{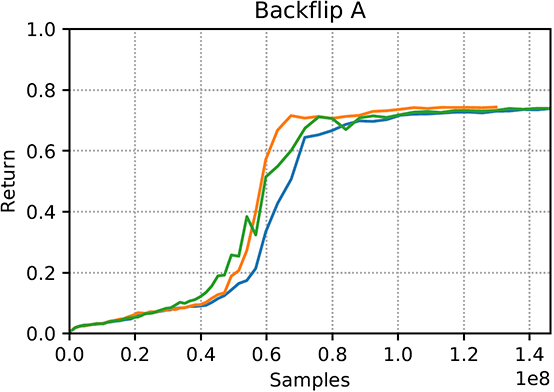}}
     \subfigure{   \includegraphics[width=0.5\columnwidth]{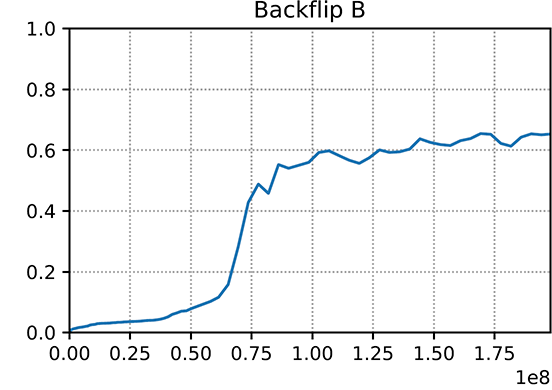}}
     \subfigure{   \includegraphics[width=0.5\columnwidth]{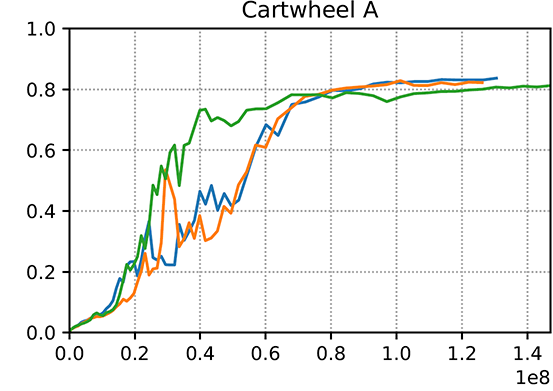}}
     \subfigure{   \includegraphics[width=0.5\columnwidth]{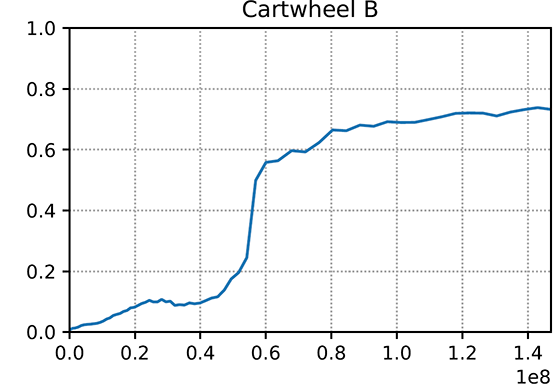}}
     \subfigure{   \includegraphics[width=0.5\columnwidth]{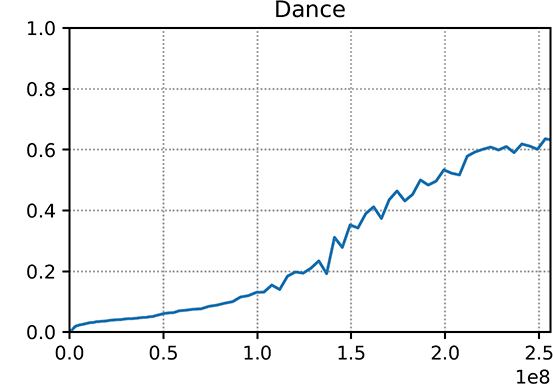}}
     \subfigure{   \includegraphics[width=0.5\columnwidth]{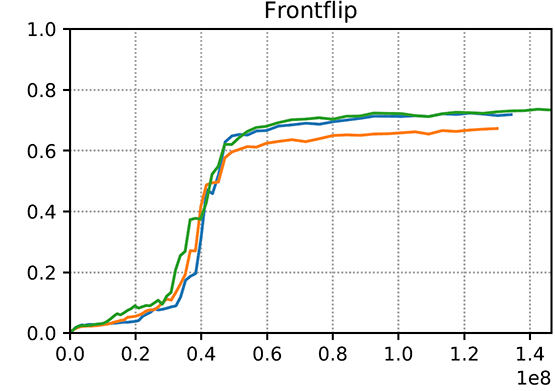}}
     \subfigure{   \includegraphics[width=0.5\columnwidth]{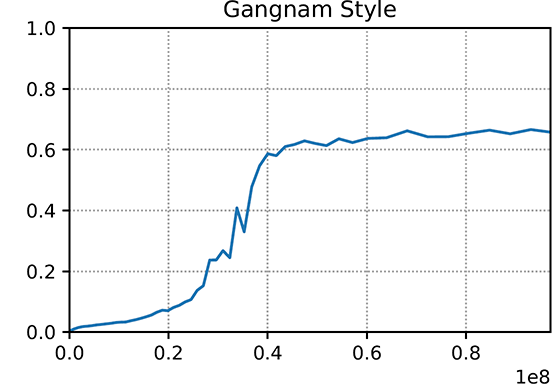}}
     \subfigure{   \includegraphics[width=0.5\columnwidth]{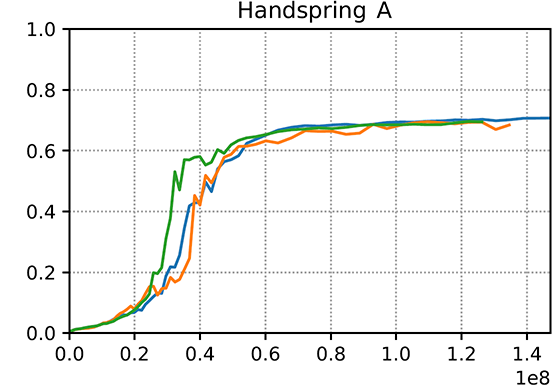}}
     \subfigure{   \includegraphics[width=0.5\columnwidth]{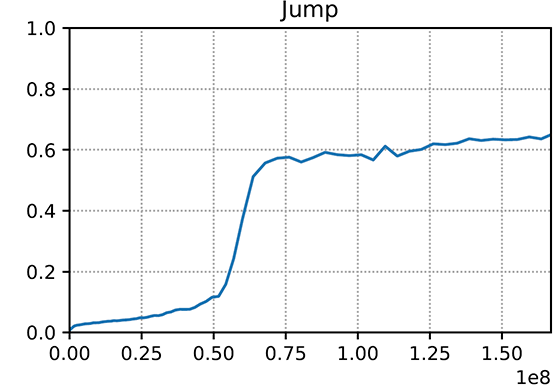}}
     \subfigure{   \includegraphics[width=0.5\columnwidth]{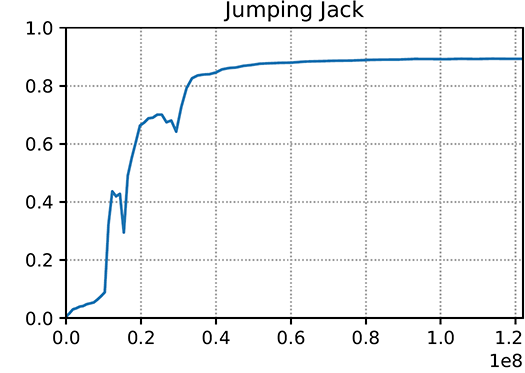}}
     \subfigure{   \includegraphics[width=0.5\columnwidth]{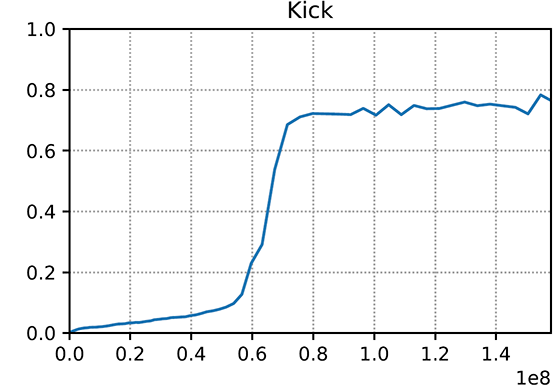}}
     \subfigure{   \includegraphics[width=0.5\columnwidth]{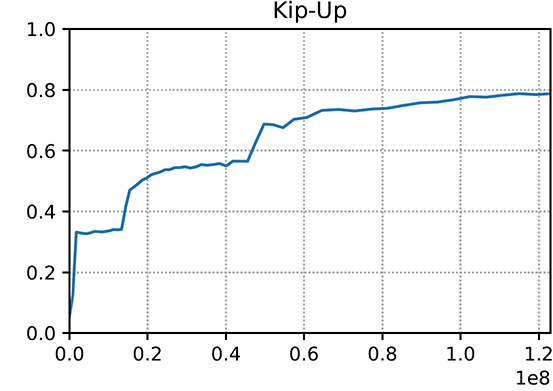}}
     \subfigure{   \includegraphics[width=0.5\columnwidth]{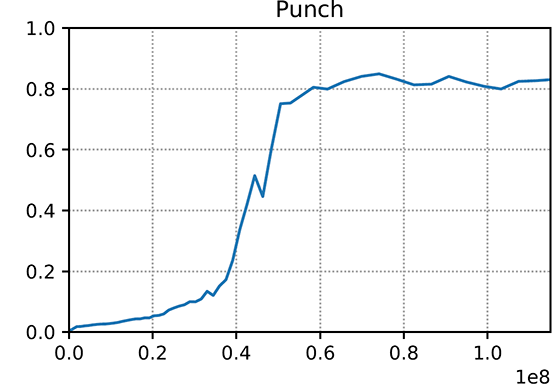}}
     \subfigure{   \includegraphics[width=0.5\columnwidth]{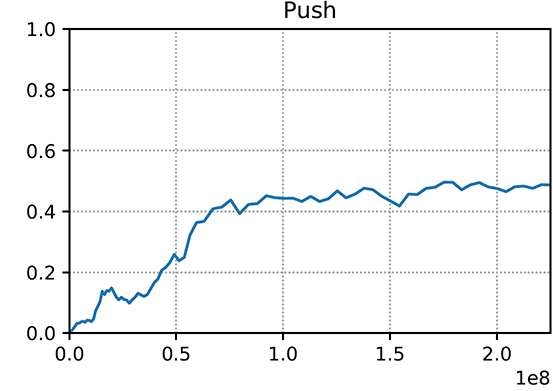}}
     \subfigure{   \includegraphics[width=0.5\columnwidth]{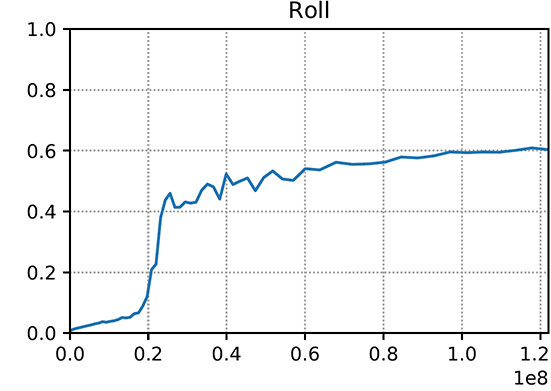}}
     \subfigure{   \includegraphics[width=0.5\columnwidth]{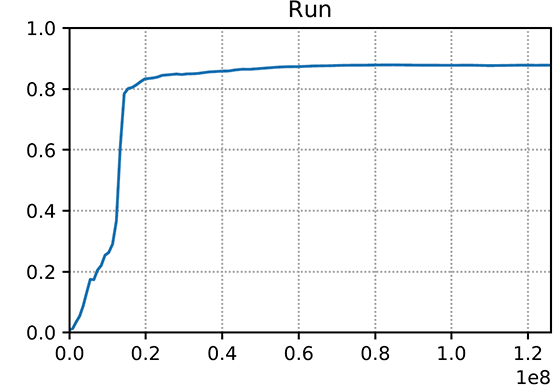}}
     \subfigure{   \includegraphics[width=0.5\columnwidth]{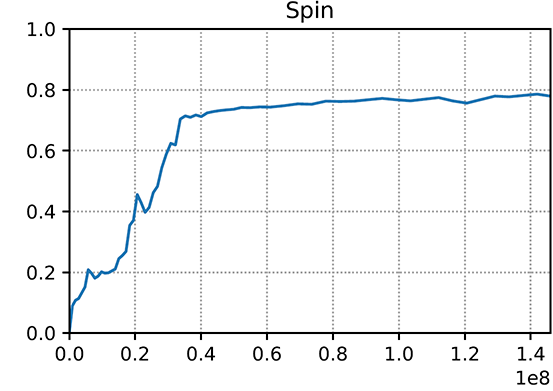}}
     \subfigure{   \includegraphics[width=0.5\columnwidth]{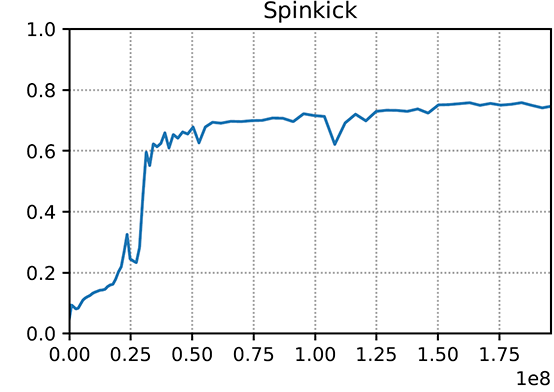}}
     \subfigure{   \includegraphics[width=0.5\columnwidth]{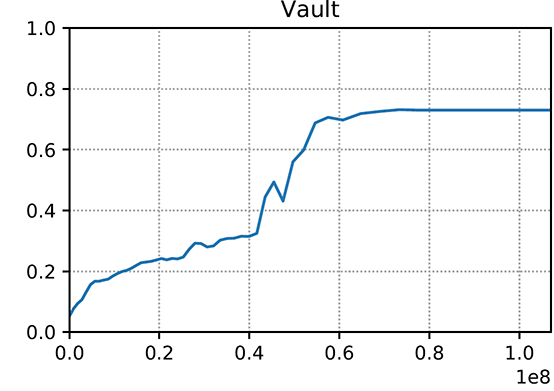}}
     \subfigure{   \includegraphics[width=0.5\columnwidth]{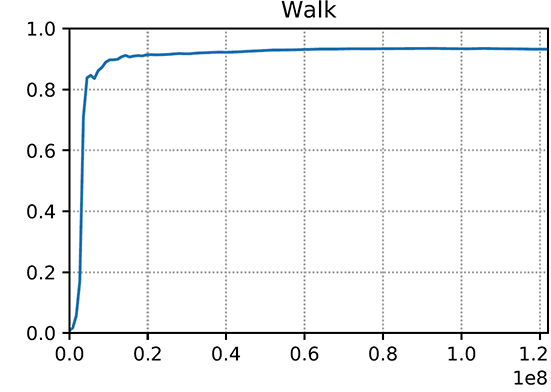}}
\caption{Learning curves of policies trained for the humanoid. Performance is recorded as the average normalized return over multiple episodes, with 0 representing the minimum possible return and 1 being the maximum.}
\label{fig:curves}
\vspace{-0.25cm}
\end{figure*}

\begin{figure*}[t]
	\centering
     \subfigure{   \includegraphics[width=0.5\columnwidth]{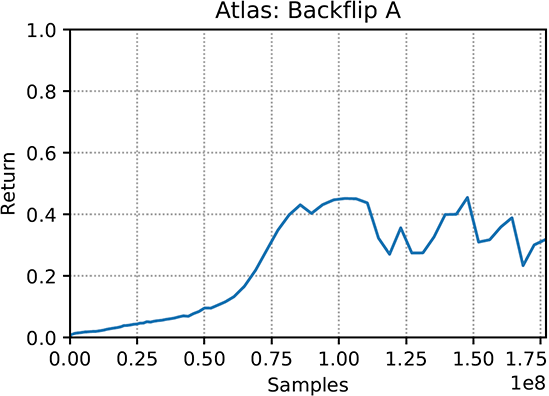}}
     \subfigure{   \includegraphics[width=0.5\columnwidth]{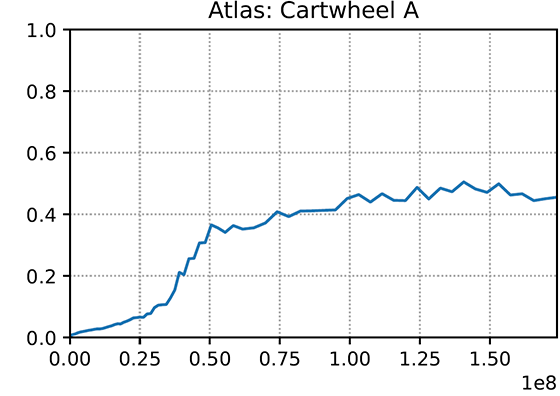}}
     \subfigure{   \includegraphics[width=0.5\columnwidth]{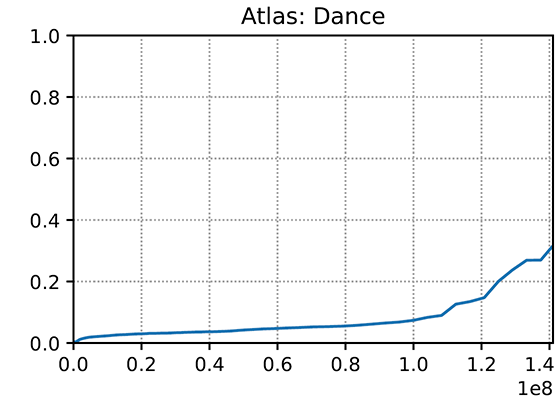}}
     \subfigure{   \includegraphics[width=0.5\columnwidth]{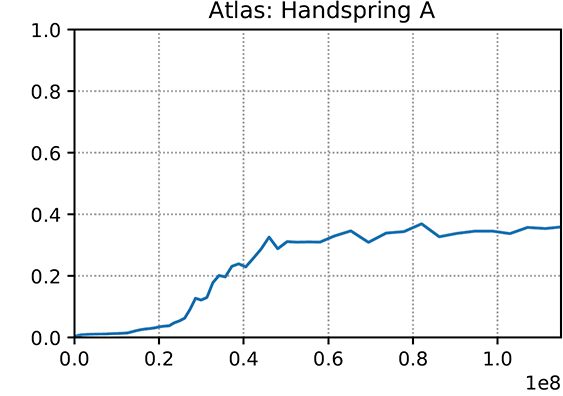}}
     \subfigure{   \includegraphics[width=0.5\columnwidth]{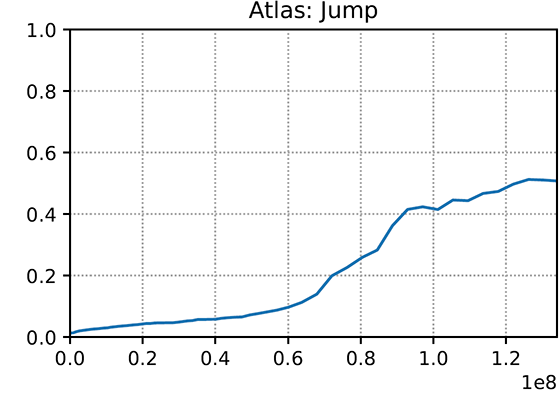}}
     \subfigure{   \includegraphics[width=0.5\columnwidth]{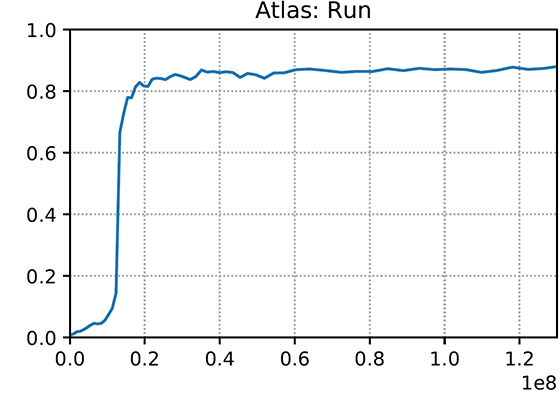}}
     \subfigure{   \includegraphics[width=0.5\columnwidth]{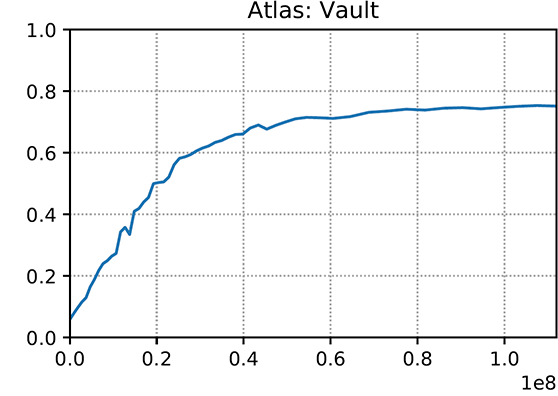}}
     \subfigure{   \includegraphics[width=0.5\columnwidth]{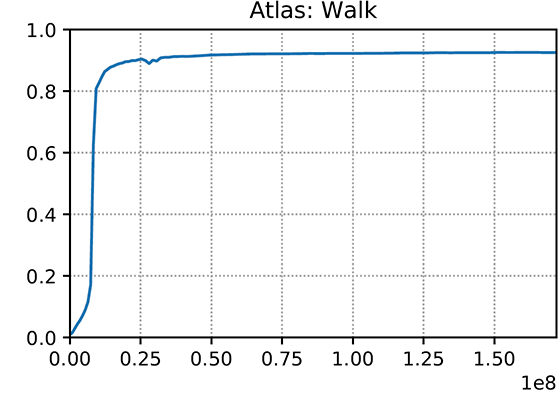}}
\caption{Learning curves of policies trained for the Atlas.}
\label{fig:curvesAtlas}
\vspace{-0.25cm}
\end{figure*}

\end{document}